\documentclass[journal]{IEEEtran}
\ifCLASSINFOpdf
\else
\fi

\usepackage{graphicx}
\usepackage{amssymb}
\usepackage{amsmath}
\usepackage{booktabs}
\usepackage{multirow}

\graphicspath{{figures/}}
\usepackage[outdir=figures_pdf/]{epstopdf}

\usepackage{xcolor}
\definecolor{ECS-Blue}{rgb}{0,0.4392,0.7529}
\definecolor{ECS-Red}{rgb}{0.7529,0,0}
\definecolor{ECS-Green}{rgb}{0.4667,0.5765,0.2353}

\usepackage[para]{threeparttable}
\makeatletter
\def\TPT@doparanotes{\par
   \prevdepth\z@ \TPT@hsize
   \TPTnoteSettings
   \parindent\z@ \pretolerance 8
   \linepenalty 200
   \renewcommand\item[1][]{\relax\ifhmode \begingroup
       \unskip
       \advance\hsize 10em % \hsize is scratch register, based on real hsize
       \penalty -45 \hskip\z@\@plus\hsize \penalty-19
       \hskip .15\hsize \penalty 9999 \hskip-.15\hsize
       \hskip .01\hsize\@plus-\hsize\@minus.01\hsize 
       \hskip 0em\@plus .3em
      \endgroup\fi
      \tnote{##1}\,\ignorespaces}%
   \let\TPToverlap\relax
   \def\endtablenotes{\par}%
}

\begin{document}

\renewcommand{\theenumi}{\alph{enumi}}
\setlength{\abovedisplayskip}{4pt plus 1pt minus 1pt}
\setlength{\belowdisplayskip}{4pt plus 0pt minus 1pt}
\setlength{\abovecaptionskip}{0pt plus 1pt minus 1pt}
\setlength{\belowcaptionskip}{0pt plus 0pt minus 1pt}
\setlength{\textfloatsep}{3pt plus 1pt minus 1pt}
\setlength{\floatsep}{3pt plus 0pt minus 1pt}
\setlength{\dbltextfloatsep}{3pt plus 1pt minus 1pt}
\setlength{\dblfloatsep}{3pt plus 0pt minus 1pt}

\title{\vspace{-0.5cm}MANTIS: A \underline{M}ixed-Sign\underline{a}l \underline{N}ear-Sensor Convolu\underline{t}ional \underline{I}mager \underline{S}oC Using Charge-Domain 4b-Weighted 5-to-84-TOPS/W MAC Operations for Feature Extraction and Region-of-Interest Detection}


\author{Martin~Lefebvre,~\IEEEmembership{Graduate Student Member,~IEEE},
        and David~Bol,~\IEEEmembership{Senior Member,~IEEE}% <-this % stops a space
\vspace{-0.75cm}
\thanks{
Received 26 June 2024; revised 5 September 2024 and 14 October 2024; accepted 18 October 2024. This article was approved by Associate Editor Yoonmyung Lee. \textit{(Corresponding author: Martin Lefebvre.)}

The authors are with the Institute of Information and Communication Technologies, Electronics and Applied Mathematics, Université catholique de Louvain, 1348 Louvain-la-Neuve, Belgium. M. Lefebvre is also with the Department of Microelectronics, Delft University of Technology, 2628 CD Delft, The Netherlands (e-mail: m.lefebvre@tudelft.nl).

Color versions of one or more figures in this article are available at https://doi.org/10.1109/JSSC.2024.3484766.

Digital Object Identifier 10.1109/JSSC.2024.3484766
}}


\markboth{IEEE Journal of Solid-State Circuits,~Vol.~xx, No.~xx, xx~2024}%
{Shell \MakeLowercase{\textit{et al.}}: Bare Demo of IEEEtran.cls for IEEE Journals}
\IEEEoverridecommandlockouts
\IEEEpubid{\begin{minipage}{\textwidth}\ \\[12pt] \begin{scriptsize}This document is the paper as accepted for publication in JSSC, the fully edited paper is available at https://ieeexplore.ieee.org/document/10750406. \copyright 2024 IEEE. Personal use of this material is permitted. Permission from IEEE must be obtained for all other uses, in any current or future media, including reprinting/republishing this material for advertising or promotional purposes, creating new collective works, for resale or redistribution to servers or lists, or reuse of any copyrighted component of this work in other works.\end{scriptsize}
\end{minipage}}
\maketitle

\begin{abstract} Recent advances in artificial intelligence have prompted the search for enhanced algorithms and hardware to support the deployment of machine learning at the edge. More specifically, in the context of the Internet of Things (IoT), vision chips must be able to fulfill tasks of low to medium complexity, such as feature extraction or region-of-interest (RoI) detection, with a sub-mW power budget imposed by the use of small batteries or energy harvesting. Mixed-signal vision chips relying on in- or near-sensor processing have emerged as an interesting candidate, thanks to their favorable tradeoff between energy efficiency (EE) and computational accuracy compared to digital systems for these specific tasks.
In this paper, we introduce a mixed-signal convolutional imager system-on-chip (SoC) codenamed MANTIS, featuring a unique combination of large 16$\times$16 4b-weighted filters, operation at multiple scales, and double sampling, well suited to the requirements of medium-complexity tasks. The main contributions are (i) circuits called DS3 units combining delta-reset sampling, image downsampling, and voltage downshifting, and (ii) charge-domain multiply-and-accumulate (MAC) operations based on switched-capacitor amplifiers and charge sharing in the capacitive DAC of the successive-approximation ADCs. MANTIS achieves peak EEs normalized to 1b operations of 4.6 and 84.1~TOPS/W at the accelerator and SoC levels, while computing feature maps with a root mean square error ranging from 3 to 11.3$\%$. It also demonstrates a face RoI detection with a false negative rate of 11.5$\%$, while discarding 81.3$\%$ of image patches and reducing the data transmitted off chip by 13$\times$ compared to the raw image.
\end{abstract}


% \fi
\IEEEpeerreviewmaketitle

\section{Introduction}
\label{sec:1_introduction}
\begin{figure}[!t]
	\centering
	\includegraphics[width=.4875\textwidth]{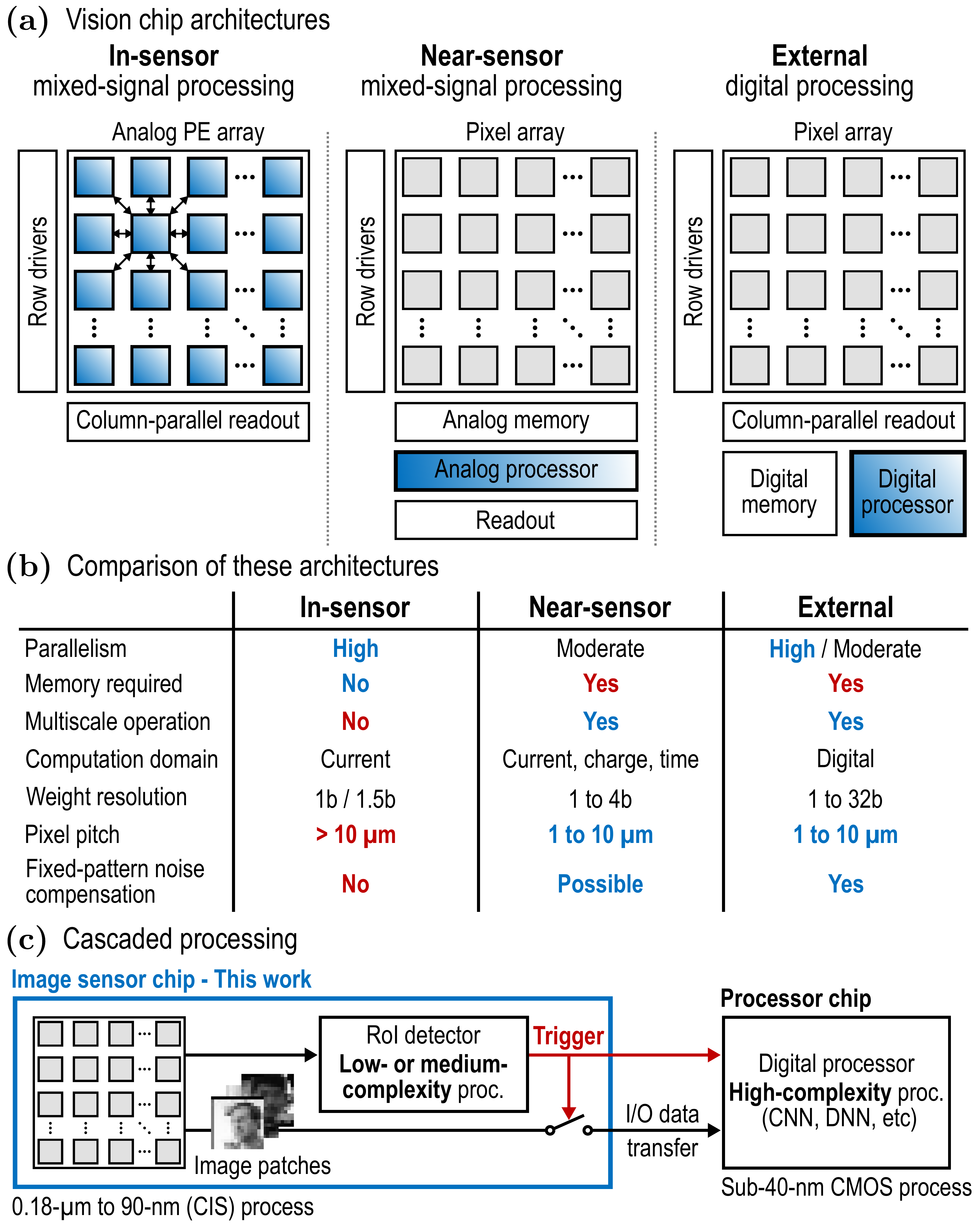}
	\caption{(a) Vision chip architectures ranging from mixed-signal processing in or near the pixel array to conventional digital processing outside of it, and (b) strengths and limitations of these architectures. (c) Envisioned system based on a cascaded processing scheme similar to \cite{Kim_2017}, in which only relevant image patches are transmitted from the image sensor to the digital processor.}
	\label{fig:1_context}
	\vspace{-0.5cm}
\end{figure}
\begin{figure}[!t]
	\centering
	\includegraphics[width=.45\textwidth]{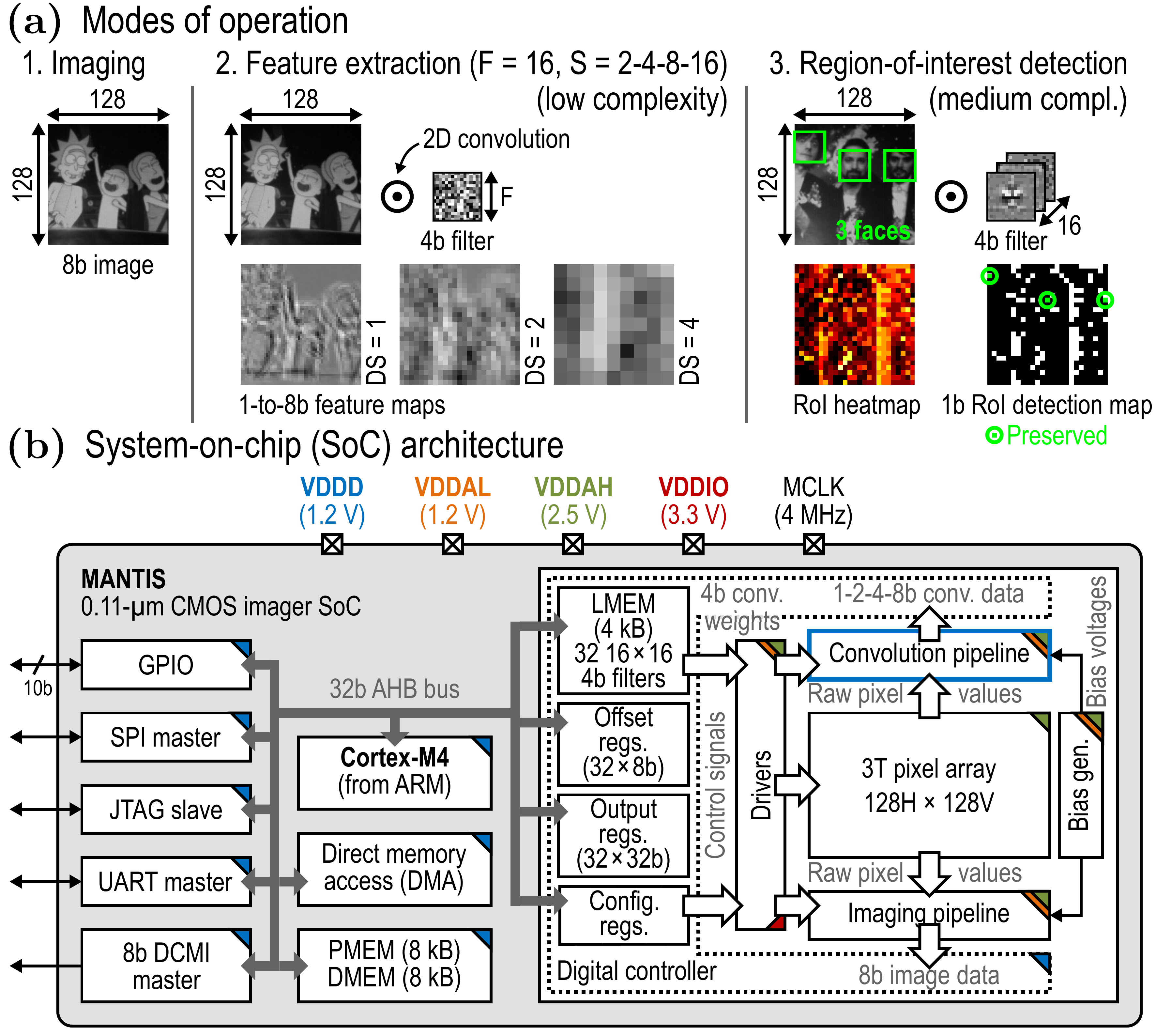}
	\caption{MANTIS CMOS imager SoC (a) modes of operation and (b) architecture, detailing the different blocks in the digital core and image sensor analog core with their respective power domains.}
	\label{fig:2_SoC_architecture}
\end{figure}

\IEEEPARstart{R}{ecent} years have seen artificial intelligence (AI) rise as a key component of numerous engineered systems, reaching an unprecedented level of pervasiveness at the applications level. Among them, the Internet of Things (IoT) has elicited a particular interest as the large amount of data generated by sensor nodes calls for the development of specialized machine learning (ML) algorithms and hardware to efficiently process data at the edge, a concept coined as edge AI or tiny ML. More specifically, in the context of vision sensors, edge devices must be able to solve vision tasks of low to medium complexity, e.g., feature extraction (FE) and region-of-interest (RoI) detection, within a \mbox{sub-mW} power budget, as IoT nodes are often supplied by limited-capacity batteries. Mixed-signal vision chips have thus emerged as a suitable candidate, since they outperform digital chips in terms of energy efficiency (EE) while maintaining a sufficient computational accuracy.
\IEEEpubidadjcol
This improved EE also stems from a reduced number of ADC conversions compared to digital implementations, leading to significant energy and area savings.\\
\indent Mixed-signal vision chip architectures can be divided into two main categories, namely in- \cite{Jendernalik_2013, Carey_2013, Xu_2022} and near-sensor \cite{Kim_2017, Bong_2018, Young_2019, Hsu_2021, Hsu_2023} vision chips, respectively implemented with analog processing elements (PEs) inside or in the periphery of the pixel array. A third category, referred to as hybrid vision chips \cite{Lefebvre_2021, Song_2021}, is not represented in Fig.~\ref{fig:1_context}(a) but simply combines elements from both categories. On the one hand, in-sensor vision chips are massively parallel and do not require any memory, be it analog or digital. However, connections between pixel-level PEs are usually local and limited to neighboring pixels, hampering the calculation of image-level features and thereby, the operation at multiple spatial scales. In addition, pixel-level PEs also lead to a relatively large pixel pitch above 10~$\mu$m. At last, in-sensor processing is often limited to low-complexity tasks as it relies on binary (1b) or ternary (1.5b) weights, and on raw \cite{Jendernalik_2013, Carey_2013, Xu_2022} or amplified \cite{Lefebvre_2021} photocurrents subject to significant fixed-pattern noise (FPN), i.e., local mismatch between pixel responses. On the other hand, near-sensor and hybrid vision chips usually present a decreased throughput compared to in-sensor ones, but are better suited to the execution of medium-complexity tasks thanks to the use of large-size 1.5b Haar-like filters \cite{Kim_2017, Bong_2018, Lefebvre_2021, Song_2021} or to an increased 4b filter weight resolution \cite{Hsu_2021, Hsu_2023}. They make use of conventional pixel structures such a three- or four-transistor (3T or 4T) active pixel sensor (APS) or pulse-width-modulated (PWM) digital pixel sensor, which are compatible with double sampling techniques to compensate FPN.
3T/4T pixels respectively use rolling and global shutters, and can either rely on a voltage-based readout with a source follower (SF), or on a time-based one with a PWM structure, allowing to reduce the supply voltage without degrading the output dynamic range. Besides, near-sensor and hybrid vision chips can operate at multiple spatial scales thanks to image downsampling (DS) \cite{Kim_2017, Bong_2018} or filter dilation \cite{Lefebvre_2021}. Finally, an analog memory generally based on capacitors (caps) is required to store a few rows of the image, ultimately leading to power and/or area overheads. Nevertheless, existing works fall short of preserving EE while simultaneously supporting medium-complexity tasks, which require sufficient computational accuracy brought by FPN-compensated inputs and increased weight resolution, as well as multiscale operation and large filters for tasks as RoI detection.\\
\indent In this work, we present a mixed-signal near-sensor convolutional imager system-on-chip (SoC) codenamed MANTIS, fabricated in United Microelectronics Corporation (UMC) \mbox{0.11-$\mu$m} CMOS technology and supporting both FE and RoI detection. It includes two main contributions providing an effective answer to the aforementioned limitations of existing vision chips. First, circuits called DS3 units combining three operations which can be abbreviated as DS, namely \underline{d}ouble \underline{s}ampling, to mitigate the impact of FPN, voltage \underline{d}own\underline{s}hifting, to reduce the voltage level from the pixel array to the convolution processor, and image \underline{d}own\underline{s}ampling, to allow for multiscale operation. Second, a mixed-signal convolution processor implementing 4b-weighted multiply-and-accumulate (MAC) operations in the charge domain, based on a modified switched-capacitor (SC) amplifier structure to compute the partial sum (psum) of a row of an image patch, and on a charge sharing operation in the capacitive digital-to-analog converter (DAC) of the following successive-approximation (SAR) analog-to-digital converter (ADC) to aggregate psums of different rows. In our vision, MANTIS would be used as the first stage of a cascaded processing system [Fig.~\ref{fig:1_context}(c)] supporting low- to medium-complexity processing tasks, while a high-complexity processing based on convolutional or deep neural networks (CNNs or DNNs) would be executed by a digital processor. The major benefits of such a system are to limit the amount of I/O data transfers from the image sensor to the digital processor, and to only dedicate energy to the processing of relevant data. This paper extends our conference paper \cite{Lefebvre_2024} by providing a more in-depth description of the circuits constituting the convolution pipeline, highlighted in Fig.~\ref{fig:2_SoC_architecture}, as well as additional experimental results. The remainder of this paper is organized as follows. First, Section~\ref{sec:2_system_on_chip_description} describes the architecture of the SoC. Then, Section~\ref{sec:3_circuit_design_and_implementation} discusses the design and implementation of the proposed mixed-signal convolution pipeline, while Section~\ref{sec:4_experimental_results} presents measurement results of the SoC. Finally, Section~\ref{sec:5_comparison_to_the_state_of_the_art} compares this work to the state of the art, and Section~\ref{sec:6_conclusion} offers some concluding remarks.
\begin{figure}[!t]
	\centering
	\includegraphics[width=.4688\textwidth]{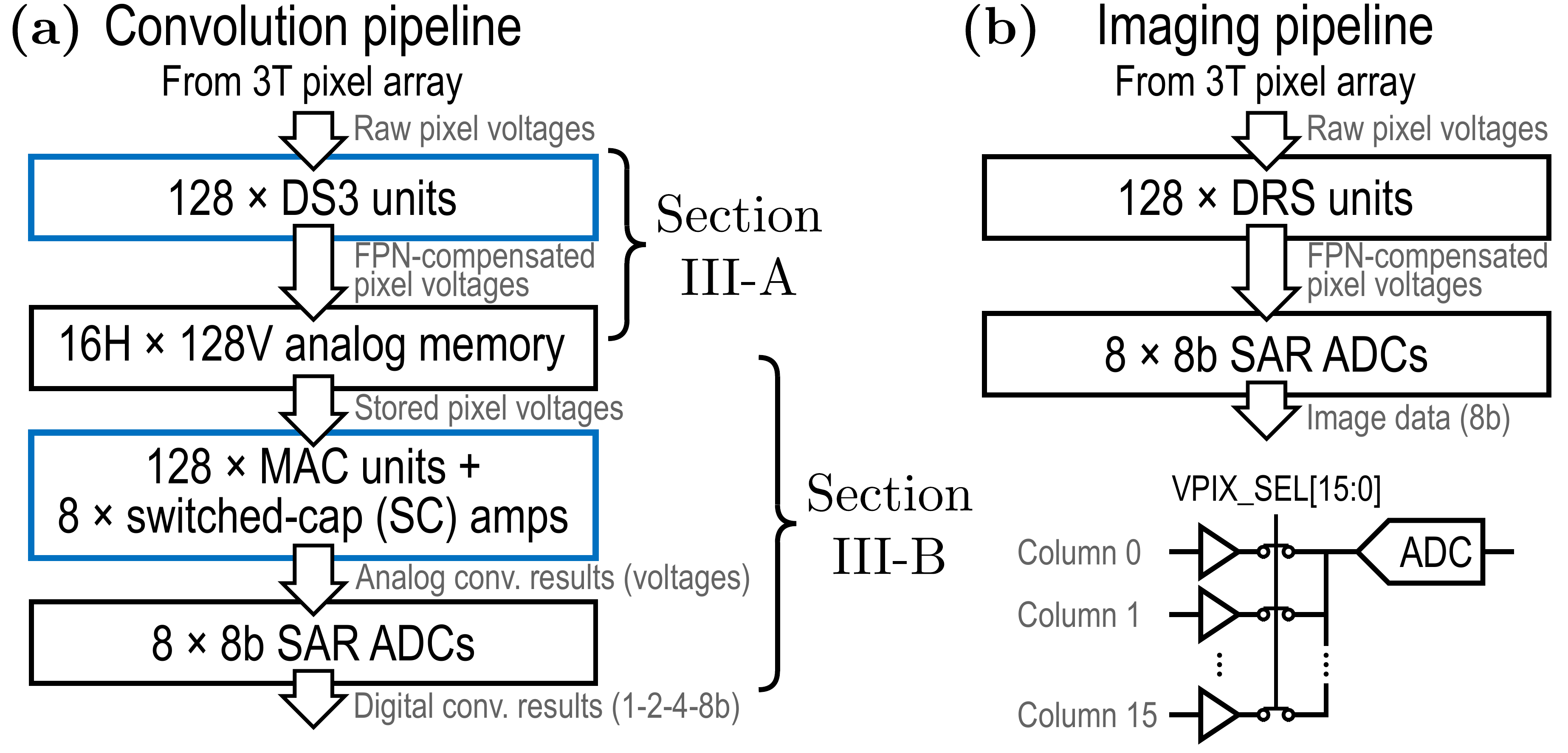}
	\caption{Block diagram of (a) the convolution and (b) the imaging pipelines.}
	\label{fig:3_pipelines}
\end{figure}

\section{System-on-Chip Description}
\label{sec:2_system_on_chip_description}

\subsection{Architecture}
\label{subsec:2B_modes_and_architecture}
This section describes the modes of operation and architecture of MANTIS CMOS imager SoC, respectively depicted in Figs.~\ref{fig:2_SoC_architecture}(a) and (b). Three modes of operations are supported. First, the imaging mode produces 8b 128$\times$128 images, which are necessary to thoroughly compare the mixed-signal on-chip execution of the convolution operations, subject to analog nonidealities, to an ideal software baseline. Second, FE can be performed using 2D convolution operations between the image and \mbox{4b-weighted} filters of fixed size F = 16. All parameters are programmable, with the filter stride S and the downsampling factor DS respectively taking any power-of-two value between 2 and 16, and between 1 to 4, and the number of filters ranging from 1 to 32. This mode generates feature maps (fmaps) with a programmable power-of-two resolution between 1 and 8 bits. Finally, an RoI detection mode supports the comparison of fmap values with a different threshold for each filter directly in the SAR ADCs. These thresholds are implemented as offsets modifying the fmap values. In this last mode of operation, 1b fmaps are created by the imager, which are subsequently combined to yield an RoI heatmap and 1b detection map, as illustrated in Fig.~\ref{fig:2_SoC_architecture}(a) for the detection of faces.\\
\begin{figure}[!t]
	\centering
	\includegraphics[width=.497\textwidth]{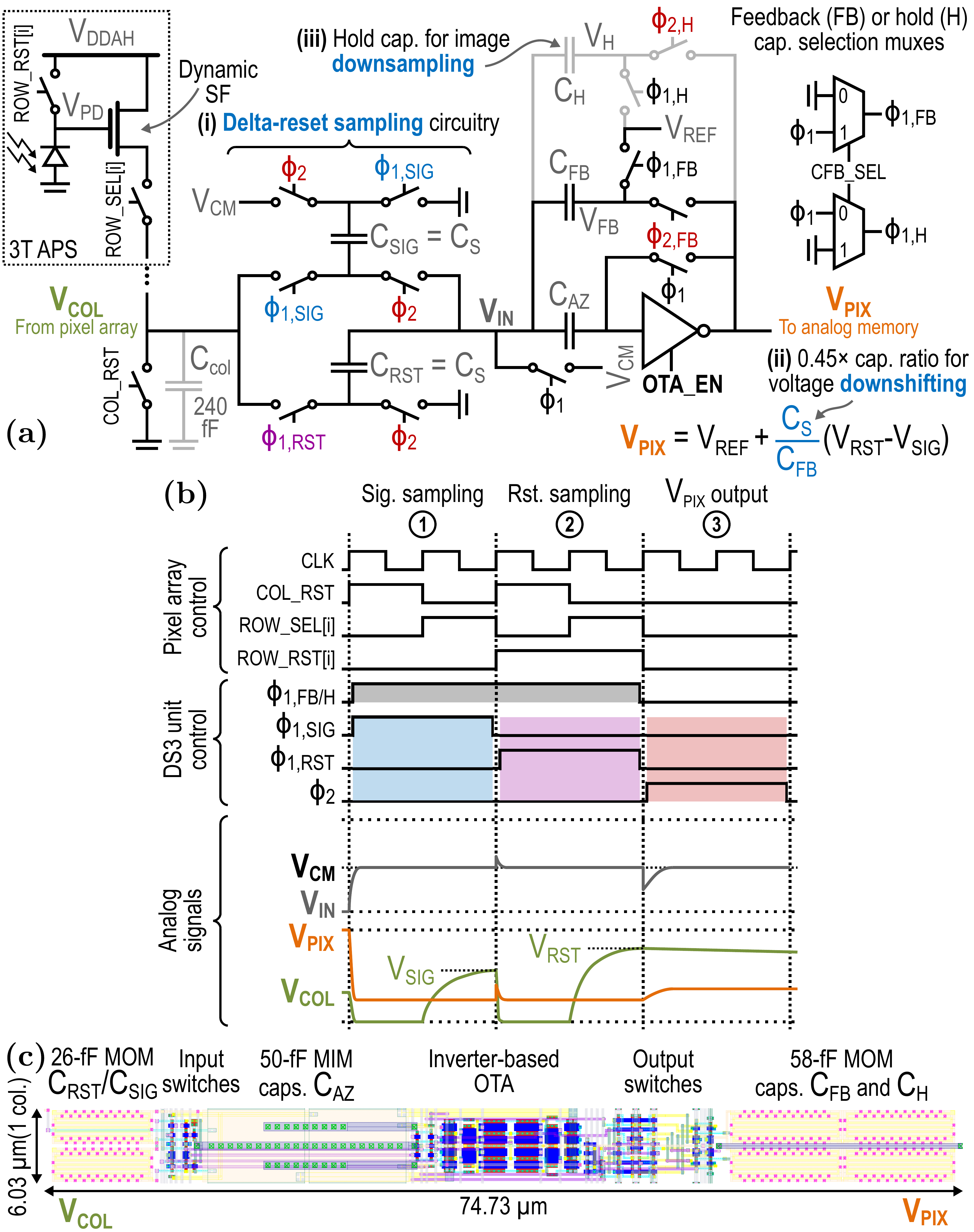}
	\caption{(a) Schematic, (b) timing diagram, and (c) 90$^\circ$-rotated layout of a single column-parallel DS3 unit. $V_\mathrm{CM}$ = 1.2~V and $V_\mathrm{REF}$ = 0.6~V in (a).}
	\label{fig:4_ds3_units_schematic}
\end{figure}
\indent Furthermore, the SoC architecture revolves around a Cortex-M4 central processing unit (CPU) from ARM, embedding a mixed-signal image sensor macro. First, regarding the \textit{digital part} of the SoC, efficient data transfers are supported by a direct memory access (DMA) peripheral allowing to move data through the advanced high-performance (AHB) bus from the imager output registers to a master digital camera interface (DCMI), which then transmits this data off chip in an 8b-parallel fashion. Moving on to the image sensor macro [Fig.~\ref{fig:2_SoC_architecture}(b) right], it includes several configuration registers for the parameters of the convolution operations discussed hereabove, among others, a \mbox{4-kB} local SRAM memory, denoted as LMEM, which can store up to 32 4b 16$\times$16 filters, and 32 8b registers for the corresponding thresholds or offsets in RoI mode. These registers impact the behavior of the digital controller piloting the analog core of the imager. Next, the \textit{analog part} of the SoC relies on a 3T 128$\times$128 pixel array with two readout pipelines: (i) a convolution pipeline supporting both FE and RoI detection modes, and (ii) an imaging pipeline used in imaging mode. It also includes a bias generation circuit used by both pipelines. The digital core is supplied at 1.2~V while the analog circuitry relies on two supplies at 1.2 and 2.5~V, respectively.

\vspace{-0.25cm}
\subsection{Convolution Pipeline}
In the convolution pipeline [Fig.~\ref{fig:3_pipelines}(a)], raw pixel voltages go through 128 column-parallel DS3 units, which zero out the FPN by a double sampling technique known as delta-reset sampling (DRS). It consists in subtracting the signal voltage of a pixel from its reset voltage, thereby suppressing the impact of local mismatch on its output. In addition, DS3 units also perform image DS to support multiscale operation. The output voltage of DS3 units are then stored in an analog memory with a capacity of 16 rows. Next, the stored pixel values are employed as inputs to 128 MAC units, connected to eight SC amplifiers computing partial convolution results or psums in the analog domain, under the form of voltages. These psums are stored in the capacitive DAC of the SAR ADCs following the SC amplifiers, before being aggregated by a charge sharing operation in the capacitive DAC and digitized to produce convolution results, with the resolution of the produced fmaps being a power of two between 1 and 8 bits.

\vspace{-0.25cm}
\subsection{Imaging Pipeline}
In the imaging pipeline [Fig.~\ref{fig:3_pipelines}(b)], DRS is used to mitigate FPN, as is done in DS3 units in the convolution pipeline. DRS units also implement voltage downshifting to adapt the \mbox{2.5-V} signals from the pixel array to the \mbox{1.2-V} input of the 8b SAR ADCs. Note that the outputs of column-parallel DRS units in 16 adjacent columns are multiplexed to a single ADC, leading to a total of eight ADCs to digitize a complete row.
\begin{figure}[!t]
	\centering
	\includegraphics[width=.497\textwidth]{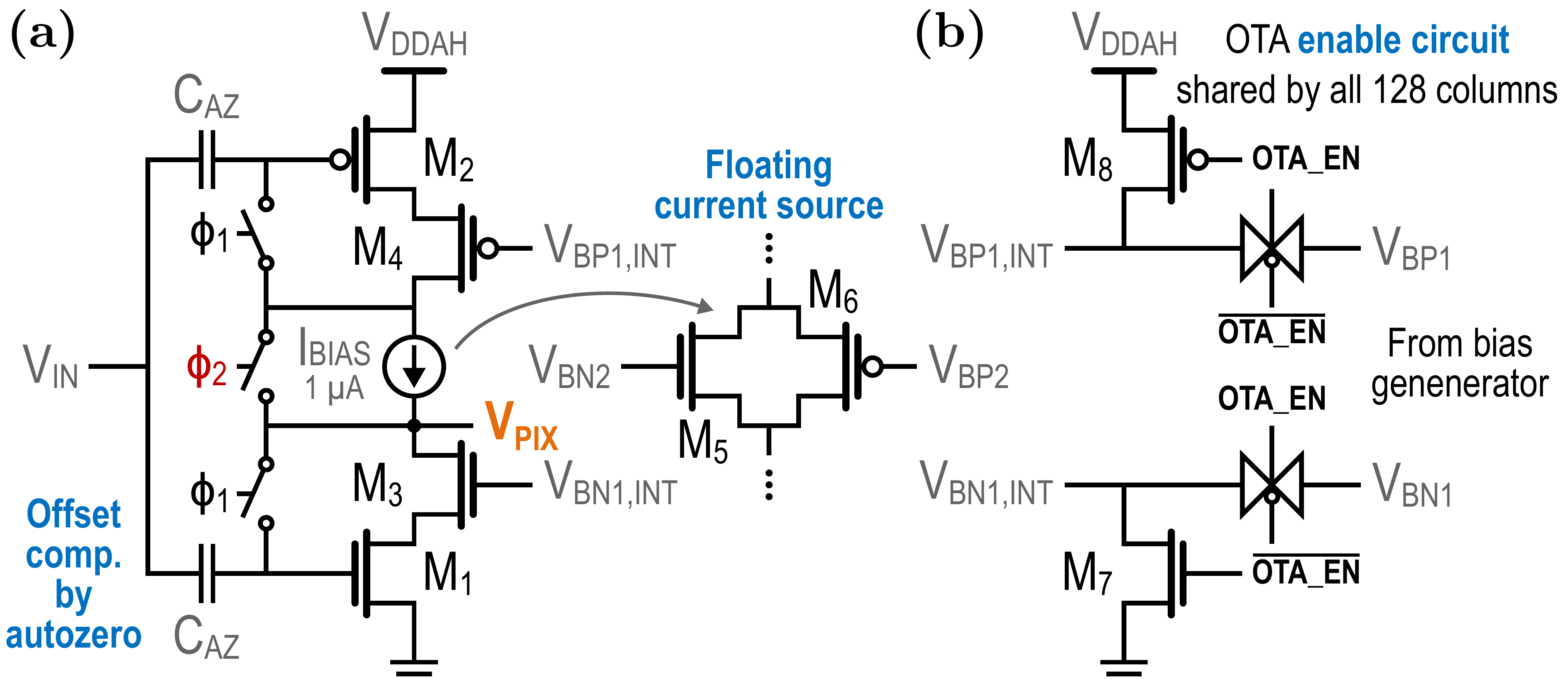}
	\caption{Schematics of (a) the inverter-based OTA proposed in \cite{Gönen_2017}, and of (b) the enable circuit shared by all 128 column-parallel DS3 units.}
	\label{fig:5_ds3_units_ota}
\end{figure}

\section{Circuit Implementation and Design of the Mixed-Signal Convolution Pipeline}
\label{sec:3_circuit_design_and_implementation}
The objective of this section is to present the circuit implementation of the mixed-signal convolution pipeline, and to provide insights regarding the design of the circuits it contains. To do so, this section is divided in two parts. Section~\ref{subsec:3A_image_readout_DS_storage} deals with the calculation of FPN-compensated downsampled pixel voltages and their storage in the analog memory, and consequently covers the DS3 units and the analog memory. Section~\ref{subsec:3B_charge_domain_MAC_operations} discusses the charge-domain MAC operations and the digitization of the convolution results, and examines the MAC units, the SC amplifiers, and the SAR ADCs.

\subsection{Image Readout, Downsampling, and Storage}
\label{subsec:3A_image_readout_DS_storage}
\subsubsection{DS3 Units for Image Readout and Downsampling}
\begin{figure}[!t]
	\centering
	\includegraphics[width=.497\textwidth]{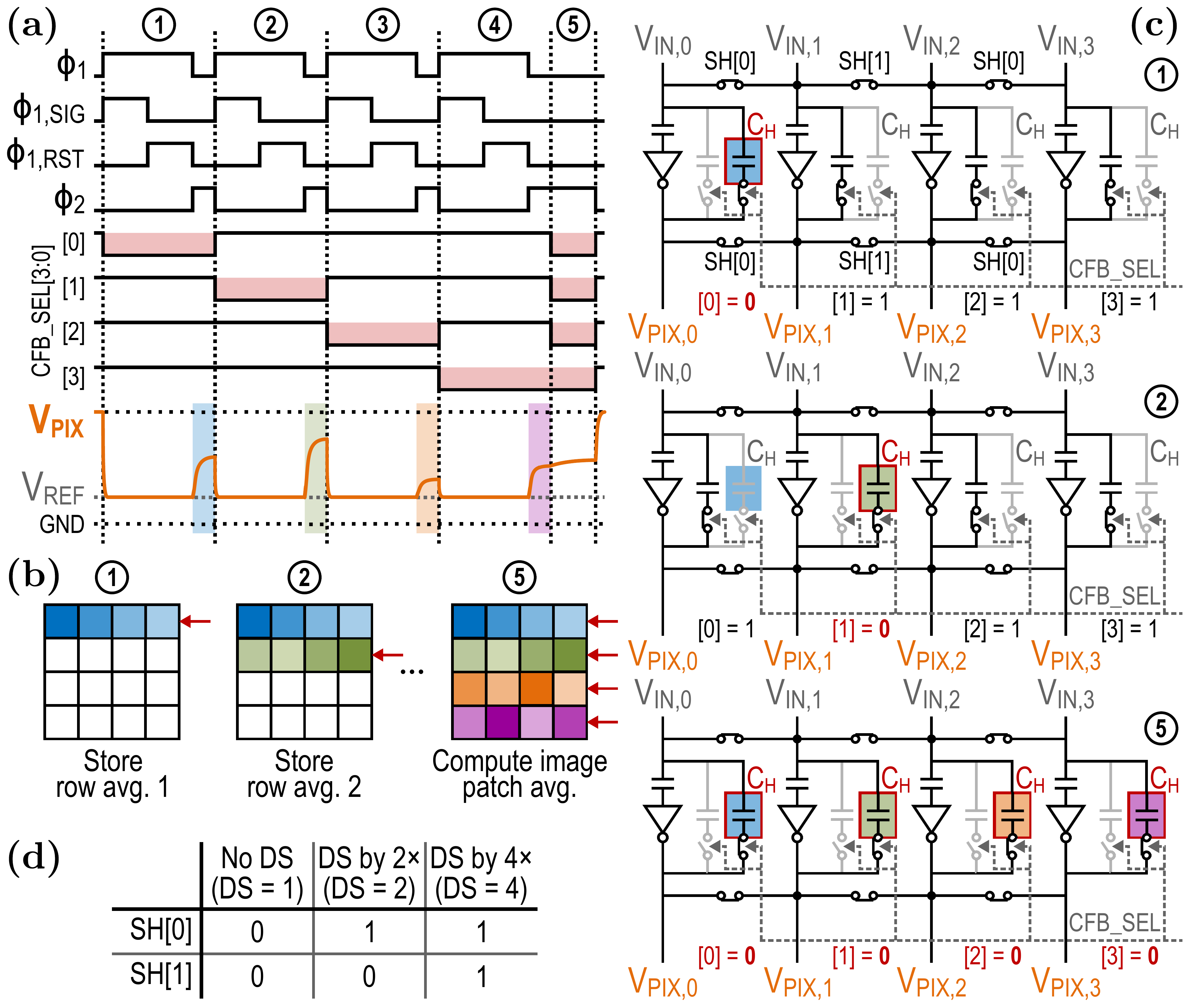}
	\caption{Illustration of image DS by 4$\times$ with (a) the timing diagram, (b) the operation principle, and (c) the schematic of four DS3 units in four neighboring columns. (d) Connection of the switches shorting the inputs and outputs of inverter-based OTAs for different DS factors.}
	\label{fig:6_ds3_units_downsampling}
\end{figure}
\begin{figure}[!t]
	\centering
	\includegraphics[width=.5\textwidth]{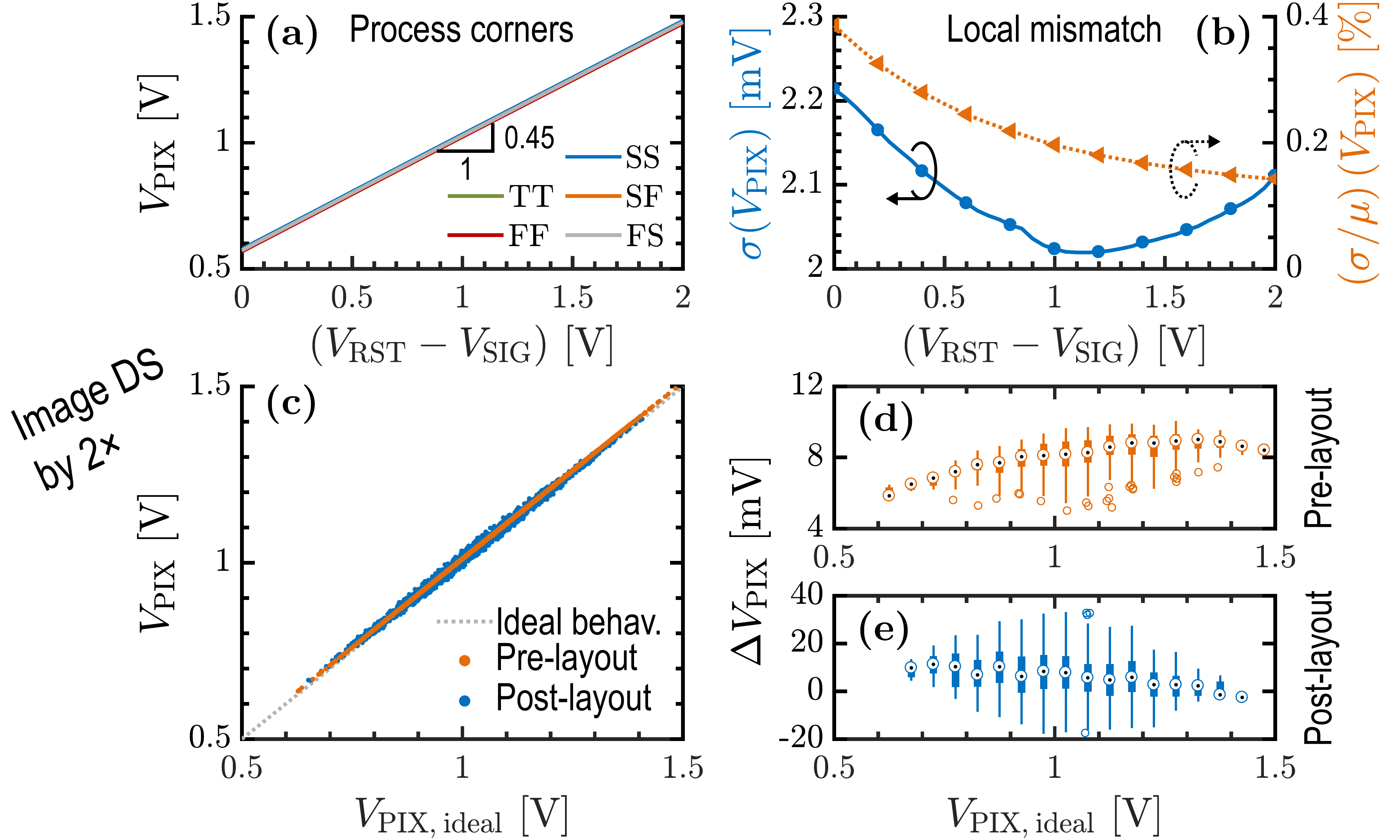}
	\caption{(a) $V_{\mathrm{PIX}}$ in process corners, and (b) variability of $V_{\mathrm{PIX}}$ for 10$^3$ MC simulations with local mismatch. Image DS by 2$\times$ for 2$\times$10$^3$ random combinations of inputs drawn from a uniform distribution, with (c) a comparison between ideal and simulated results, and distributions of the error $\Delta V_{\mathrm{PIX}} = (V_{\mathrm{PIX}}-V_{\mathrm{PIX,\:ideal}})$ in (d) pre- and (e) post-layout simulations. All figures correspond to the TT 25~$^\circ$C corner, except for (a) which covers all five process corners.}
	\label{fig:7_ds3_units_charac}
\end{figure}
Fig.~\ref{fig:4_ds3_units_schematic} illustrates the operation of a column-parallel DS3 unit to read a single pixel value while performing DRS and voltage downshifting. This operation consists of three steps [Fig.~\ref{fig:4_ds3_units_schematic}(b)]. In step $\raisebox{.5pt}{\textcircled{\raisebox{-.9pt} {1}}}$, the signal coming from the 3T APS, resulting from the discharge of the internal pixel node $V_\mathrm{PD}$ by the photocurrent during the exposure time, is read on the column voltage $V_\mathrm{COL}$. To do so, we rely on the partial settling or dynamic SF readout from \cite{Park_2020}, which consists in resetting $V_\mathrm{COL}$ to ground using the \texttt{COL\_RST} switch, before enabling the SF during a finite amount of time (\texttt{ROW\_SEL[i]} = 1). This readout is more energy-efficient than the conventional one using a current source at the bottom of the column, as it eliminates any static current consumption. It also presents an optimal settling time \cite{Park_2020} which minimizes the variability of $(V_\mathrm{RST}-V_\mathrm{SIG})$, that we find to be 0.5~$\mu$s in our design. At the end of step $\raisebox{.5pt}{\textcircled{\raisebox{-.9pt} {1}}}$, the signal value has been sampled on the \mbox{26-fF} MOM cap $C_\mathrm{SIG}$. Then, in step $\raisebox{.5pt}{\textcircled{\raisebox{-.9pt} {2}}}$, the pixel is reset and the resulting value is sampled on $C_\mathrm{RST}$, whose capacitance is the same as $C_\mathrm{SIG}$. Finally, during step $\raisebox{.5pt}{\textcircled{\raisebox{-.9pt} {3}}}$, these two caps are connected with opposite polarities, and their charges are dumped on a \mbox{58-fF} MOM feedback cap $C_\mathrm{FB}$, resulting in a voltage $V_\mathrm{PIX}$ in which $(V_\mathrm{RST}-V_\mathrm{SIG})$ is multiplied by the capacitance ratio $C_\mathrm{S}/C_\mathrm{FB}$ = 0.45. Thus, the operations of (i) DRS and (ii) voltage downshifting are realized. Moreover, in the schematic depicted in Fig.~\ref{fig:4_ds3_units_schematic}(a), switches connected to ground or $V_\mathrm{DD}$ are respectively implemented with single nMOS or pMOS, while other switches are transmission gates (TGs). They rely on \mbox{3.3-V} I/O transistors to withstand the \mbox{2.5-V} supply, with $W$ = 0.25~$\mu$m and $L$ = $L_\mathrm{min}$ = 0.34~$\mu$m, except for the TGs connected to $V_\mathrm{REF}$ and $V_\mathrm{CM}$ for which $L$ = 0.68~$\mu$m to reduce leakage. Regarding the capacitors, they are chosen to ensure that local mismatch, noise, and layout parasitics have a minimal impact on the circuit behavior, but they could be downsized as long as the uncertainty and voltage attenuation remain within the specifications of the target algorithm. Further reduction could be achieved by accounting for these nonidealities in the training algorithm, as is done in \cite{Kneip_2023} for in-memory computing (IMC). A similar design choice is made for the other mixed-signal circuits in this work.\\
\indent Besides, to compensate for the offset of the inverter-based operational transconductance amplifier (OTA) \cite{Gönen_2017}, it is put in autozero (AZ) during steps $\raisebox{.5pt}{\textcircled{\raisebox{-.9pt} {1}}}$ and $\raisebox{.5pt}{\textcircled{\raisebox{-.9pt} {2}}}$. This corresponds to sampling on the two \mbox{50-fF} MIM caps $C_\mathrm{AZ}$ the difference between the common-mode voltage $V_\mathrm{CM}$ and the $V_\mathrm{GS}$ of $M_\mathrm{1-2}$ with a fixed \mbox{1-$\mu$A} bias current, imposed by the floating current source formed by $M_\mathrm{5-6}$ [Fig.~\ref{fig:5_ds3_units_ota}(a)]. Moreover, a key feature to reach a high EE is the enable circuit shared by all amplifiers [Fig.~\ref{fig:5_ds3_units_ota}(b)], implemented by respectively clamping bias voltages $V_\mathrm{BN1,INT}$ and $V_\mathrm{BP1,INT}$ to ground and $V_\mathrm{DDAH}$, and thus allowing to duty cycle DS3 units to save power.\\
\indent Next, the image DS relies on a principle proposed in \cite{Young_2019} and represented for a DS by 4$\times$ in Fig.~\ref{fig:6_ds3_units_downsampling}. As illustrated in Fig.~\ref{fig:6_ds3_units_downsampling}(c), the inputs and outputs of the OTAs in four adjacent columns are shorted together by switches whose configuration depends on the DS factor [Fig.~\ref{fig:6_ds3_units_downsampling}(d)]. The average of each row of a 4$\times$4 image patch is computed and stored in the hold cap $C_\mathrm{H}$ of one of the four columns, as shown in steps  $\raisebox{.5pt}{\textcircled{\raisebox{-.9pt} {1}}}$ to $\raisebox{.5pt}{\textcircled{\raisebox{-.9pt} {4}}}$ [Figs.~\ref{fig:4_ds3_units_schematic}(a) to (c)]. Once all row averages have been computed, all $C_\mathrm{H}$ caps are simultaneously connected during step $\raisebox{.5pt}{\textcircled{\raisebox{-.9pt} {5}}}$, and the resulting voltage is the average of row averages, or in other words, the average of the image patch. The proposed DS3 unit fits into the \mbox{6.03-$\mu$m} pixel pitch and occupies 74.73~$\mu$m in height [Fig.~\ref{fig:4_ds3_units_schematic}(c)], a dimension which could be further reduced if transistors could be placed below MIM caps $C_\mathrm{AZ}$.\\
\begin{figure}[!t]
	\centering
	\includegraphics[width=.5\textwidth]{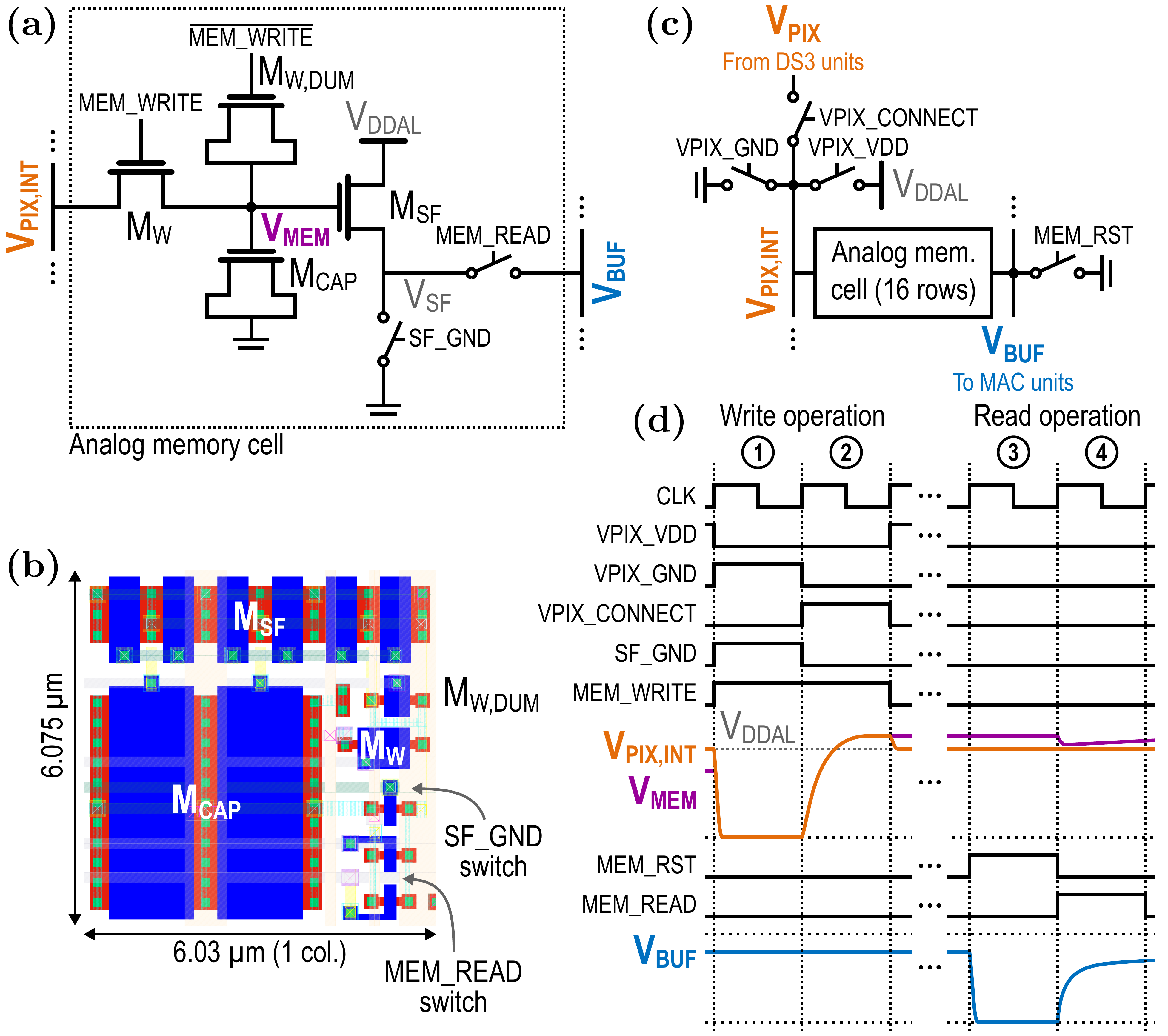}
	\caption{(a) Schematic and (b) layout of an analog memory cell. (c) Connections at the input and output of a column of 16 row of memory cells and (d) timing diagram for write and read operations.}
	\label{fig:8_analog_mem_schematic}
\end{figure}
\begin{figure}[!t]
	\centering
	\includegraphics[width=.45\textwidth]{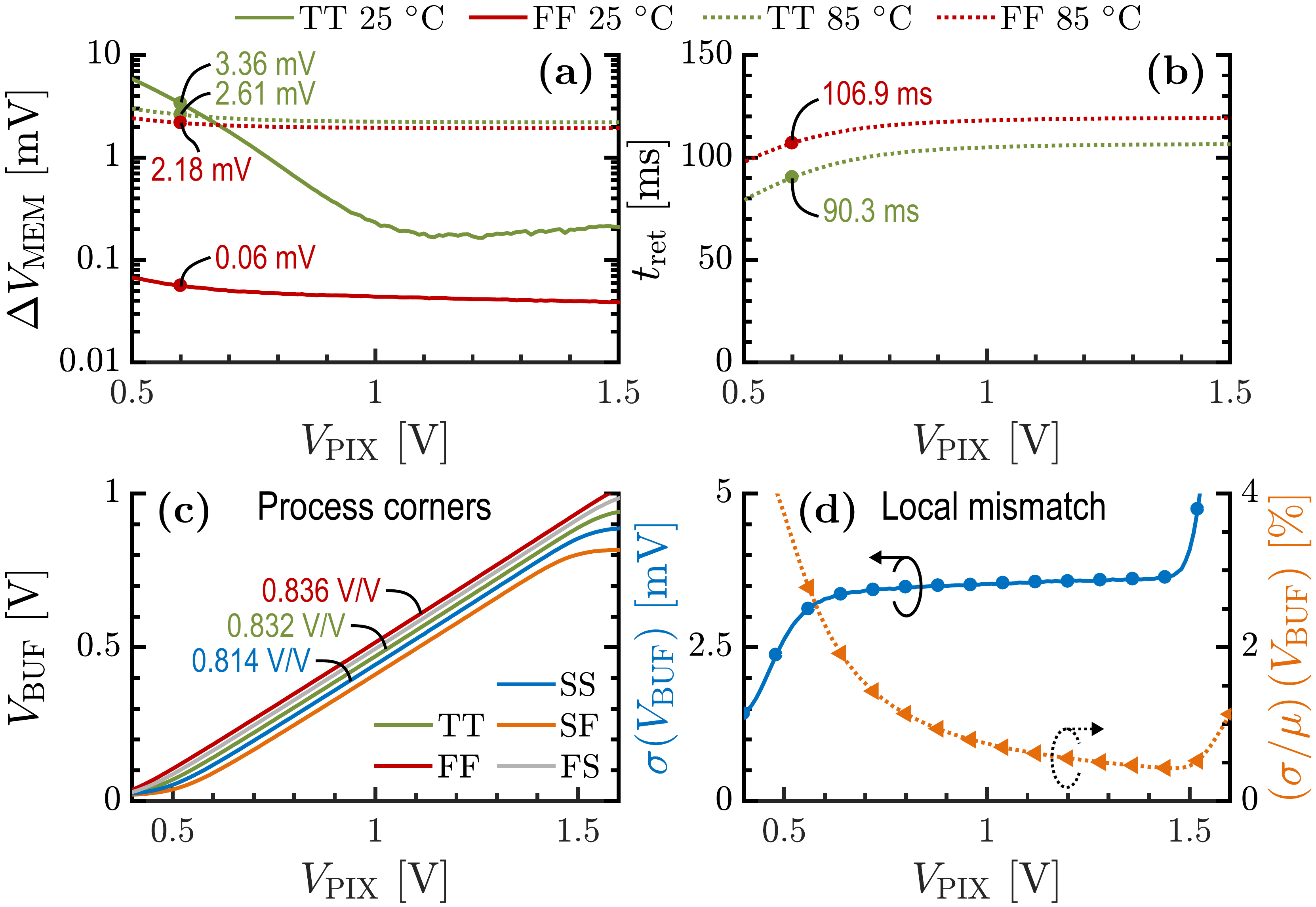}
	\caption{(a) Voltage change of $V_{\mathrm{MEM}}$ after 100~ms in TT and FF, and at 25~$^\circ$C and 85~$^\circ$C. (b) Retention time $t_\mathrm{ret}$ in TT 85~$^\circ$C and FF 85~$^\circ$C (worst case). $t_\mathrm{ret}$ is defined as the time at which the initial voltage has changed by more than 2.35~mV, corresponding to half an LSB for a 1.2-V 8b ADC. At 25~$^\circ$C, (c) transfer function from $V_\mathrm{PIX}$ to $V_\mathrm{BUF}$ in process corners and (d) variability of $V_\mathrm{BUF}$ for 10$^3$ MC simulations with local mismatch.}
	\label{fig:9_analog_mem_charac}
\end{figure}
\indent This circuit is robust to process, voltage, and temperature (PVT) variations due to its ratiometric nature, as long as the OTA is designed to operate in the relevant corners. Thus, this article does not aim at providing an exhaustive characterization of the proposed circuits in PVT corners, but focuses on their main performance and limitations. Post-layout simulations in Fig.~\ref{fig:7_ds3_units_charac} confirm the independence with respect to process [Fig.~\ref{fig:7_ds3_units_charac}(a)] and show that the $\sigma$ and $\sigma/\mu$ of $V_\mathrm{PIX}$ due to local mismatch for a single DS3 unit are respectively below 2.2~mV and 0.4$\%$ across the input range. Regarding the output voltage noise, a theoretical expression is given by $\overline{v_\mathrm{n}} = \frac{C_\mathrm{S}}{C_\mathrm{FB}} \sqrt{\frac{2kT}{C_\mathrm{S}}}$, where $k$ is Boltzmann's constant and $T$ the absolute temperature. At 25~$^\circ$C, $\overline{v_\mathrm{n}}$ = 0.25~mV and its impact is significantly lower than that of mismatch. In addition, the performance of DS is evaluated for DS by 2$\times$ in Figs.~\ref{fig:7_ds3_units_charac}(c) to (e), for 2$\times$10$^3$ combinations of input voltages drawn from a uniform distribution. The $\sigma$ of the error $\Delta V_\mathrm{PIX}$ increases from less than 1~mV pre-layout to approximately 10~mV post-layout, as highlighted in Figs.~\ref{fig:7_ds3_units_charac}(d) and (e), due to capacitive coupling between nodes $V_\mathrm{IN}$, $V_\mathrm{PIX}$ and $V_\mathrm{H}$ [Fig.~\ref{fig:4_ds3_units_schematic}(a)] which could be reduced by investing more effort into the layout.

\subsubsection{Analog Memory for Image Storage}
The schematic and operation of the analog memory are described in Fig.~\ref{fig:8_analog_mem_schematic}. A memory cell with a structure close to \cite{Kim_2017, Bong_2018} [Fig.~\ref{fig:8_analog_mem_schematic}(a)] consists of a \mbox{32-fF} MOS cap $M_\mathrm{CAP}$, an access transistor $M_\mathrm{W}$ with a dummy transistor $M_\mathrm{W,DUM}$ with half the length, to compensate for the charge injection of $M_\mathrm{W}$, and an SF $M_\mathrm{SF}$ employed in a dynamic fashion for its reduced mismatch of $V_\mathrm{BUF}$ and decreased static power consumption, similar to the pixel readout in Section~\ref{subsec:3A_image_readout_DS_storage}. This memory cell occupies a silicon area of 6.03~$\mu$m $\times$ 6.075~$\mu$m [Fig.~\ref{fig:8_analog_mem_schematic}(b)], close to that of a pixel. To read or write a cell located within a column of the analog memory [Figs.~\ref{fig:8_analog_mem_schematic}(c) and (d)], several switches are used to connect the column internal voltage $V_\mathrm{PIX,INT}$ to the output of the DS3 units, or to ground/$V_\mathrm{DDAL}$.
\begin{figure}[!t]
	\centering
	\includegraphics[width=.5\textwidth]{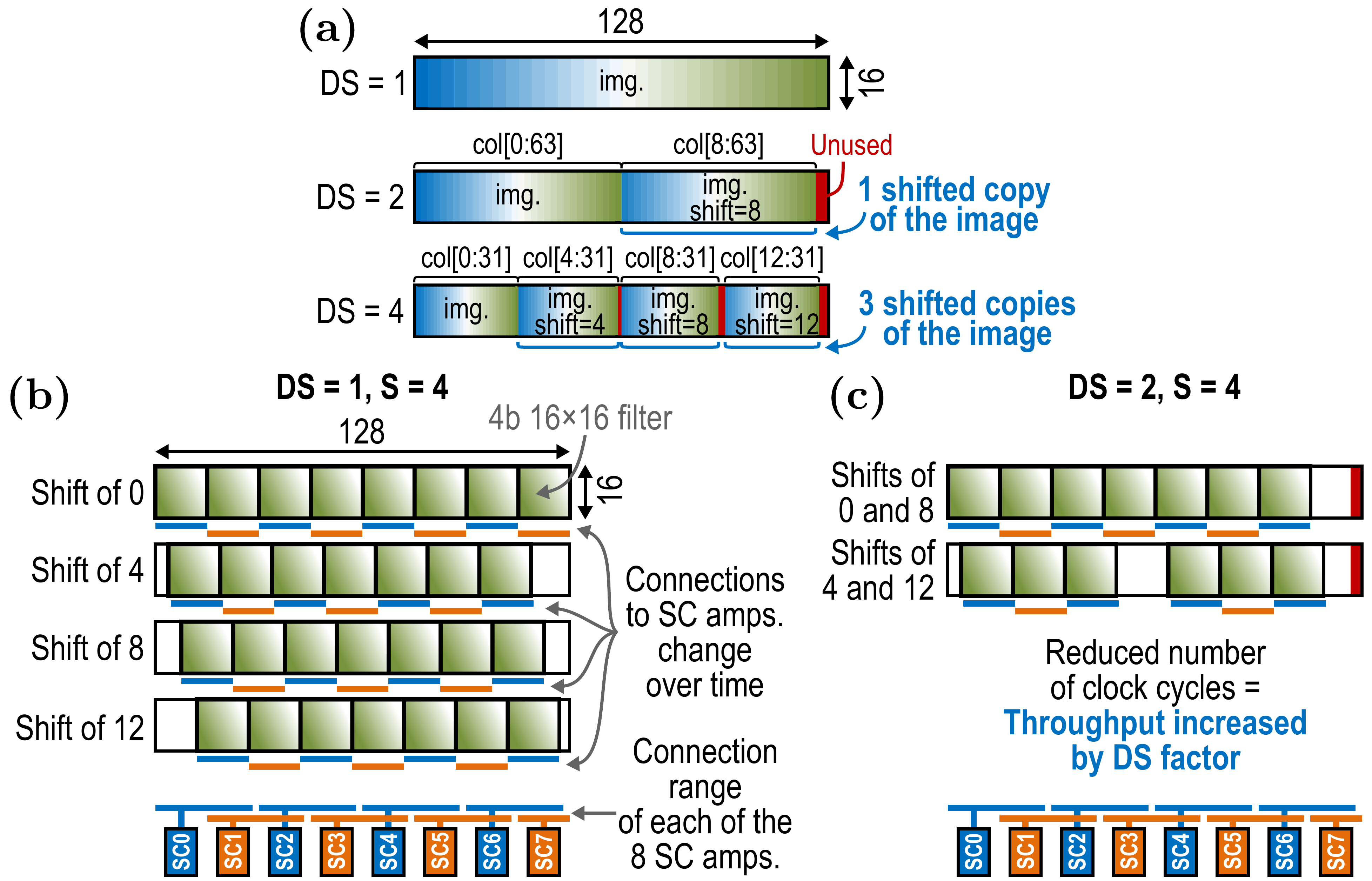}
	\caption{(a) Storage pattern of the image into the analog memory for different DS factors. Filter striding for a convolution operation with a stride S = 4 (b) without image DS (DS = 1) and (c) with image DS by 2$\times$ (DS = 2).}
	\label{fig:10_mac_units_operation_principle}
\end{figure}
\begin{figure*}[!t]
	\centering
	\includegraphics[width=.926\textwidth]{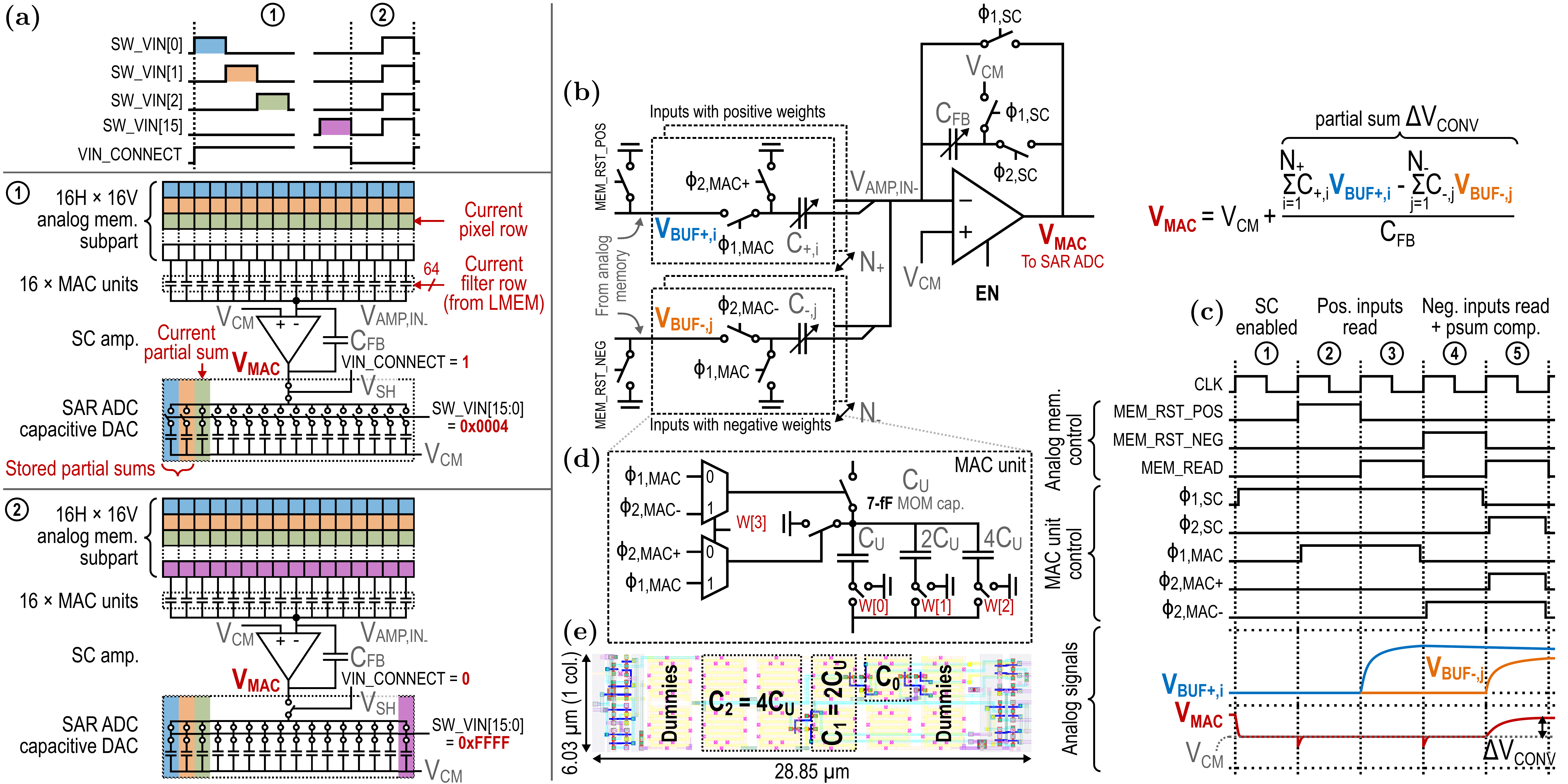}
	\caption{(a) Principle of the convolution operation between a 16$\times$16 image patch and a 4b 16$\times$16 filter. (b) Schematic of the SC amplifier realizing the MAC operations between a single row of the image patch and filter, with (c) the corresponding timing diagram. The common-mode voltage $V_\mathrm{CM}$ is equal to $V_\mathrm{DDAL}/2$ = 0.6~V. (d) Detailed schematic and (e) 90$^\circ$-rotated layout of one of the 16 MAC units connected to the SC amplifier.}
	\label{fig:11_mac_units_schematic}
\end{figure*}
During a write operation, in step $\raisebox{.5pt}{\textcircled{\raisebox{-.9pt} {1}}}$, $V_\mathrm{PIX,INT}$, $V_\mathrm{MEM}$ and $V_\mathrm{SF}$ are grounded to overwrite the memory cell content without any impact from previously stored values. Then, in step $\raisebox{.5pt}{\textcircled{\raisebox{-.9pt} {2}}}$, $V_\mathrm{MEM}$ is driven to $V_\mathrm{PIX}$ by the DS3 unit connected to the column, before disconnecting $V_\mathrm{MEM}$ from $V_\mathrm{PIX,INT}$. During a read operation, $V_\mathrm{BUF}$ is first reset to ground in step $\raisebox{.5pt}{\textcircled{\raisebox{-.9pt} {3}}}$, before reading the memory cell by partial settling in step $\raisebox{.5pt}{\textcircled{\raisebox{-.9pt} {4}}}$.\\
\indent When the memory is not written or read, the retention of the memory cells needs to be maximized in the worst-case corner, here FF 85~$^\circ$C. To do so, we minimize the leakage of the access transistor $M_\mathrm{W}$ by implementing it with a \mbox{3.3-V} I/O nMOS with $W$ = 0.18~$\mu$m and $L$ = 1~$\mu$m, and additionally, by driving $V_\mathrm{PIX,INT}$ to $V_\mathrm{DDAL}$ = 1.2~V in retention to limit the $V_\mathrm{DS}$ of $M_\mathrm{W}$ and further reduce the leakage, given that $V_\mathrm{PIX}$ approximately ranges from 0.6 to 1.5~V [Fig.~\ref{fig:7_ds3_units_charac}(a)]. Continuing with the post-layout characterization of the memory, Fig.~\ref{fig:9_analog_mem_charac}(a) highlights that the typical voltage change of the stored voltage after 100~ms is respectively 2.61 and 2.18~mV in the TT and FF process corners at 85~$^\circ$C, while Fig.~\ref{fig:9_analog_mem_charac}(b) indicates retention times of 90.3 and 106.9~ms in the same conditions, the retention time being defined as a change of $\pm$LSB/2 with respect to the initially stored voltage, i.e., $\pm$2.35~mV for a \mbox{1.2-V} supply and an 8b resolution. In addition, in Fig.~\ref{fig:9_analog_mem_charac}(a), the linear increase of $\Delta V_\mathrm{MEM}$ in TT 25~$^\circ$C for $V_\mathrm{PIX} <$ 1~V can be explained by a slow transient of $V_\mathrm{SF}$ lightly affecting $V_\mathrm{MEM}$ through capacitive coupling. In Fig.~\ref{fig:9_analog_mem_charac}(c), we observe that the transfer function of the SF has a slope $A_\mathrm{SF}$ below 1~V/V due to the body effect resulting from $M_\mathrm{SF}$'s body being grounded, and that it is impacted by variations of $M_\mathrm{SF}$'s threshold voltage in process corners even though the slope remains around 0.83~V/V. Finally, $V_\mathrm{BUF}$ has a $\sigma$ around 3.5~mV in the usable part of the input range, corresponding to a maximum $\sigma/\mu$ of 2.3$\%$ for $V_\mathrm{PIX}$ = 0.6~V, while in comparison, the output noise $\overline{v_\mathrm{n}} = A_\mathrm{SF} \sqrt{\frac{kT}{C_\mathrm{MEM}}}$ equal to 0.3~mV at 25~$^\circ$C is negligible. Future designs could compensate the SF mismatch by making use of an OTA-based feedback loop to write the analog memory, as proposed by Seo et al. in \cite{Seo_2023}.

\subsection{Charge-Domain Multiply-and-Accumulate Operations}
\label{subsec:3B_charge_domain_MAC_operations}
\subsubsection{Operation Principle}
When no DS is applied to the image, the columns of the pixel array match that of the analog memory in a one-to-one fashion, and the whole width of the analog memory is used to store the image [Fig.~\ref{fig:10_mac_units_operation_principle}(a)]. The convolution operation is thus performed between several replicas of the 4b 16$\times$16 filter and different image patches without overlap [Fig.~\ref{fig:10_mac_units_operation_principle}(b)]. As the filter is shifted to the right, the connections between the analog memory and the eight SC amplifiers are modified over time to follow the movement of the filter. However, when a DS by 2$\times$ is applied, the image is only 64-columns wide, so the first half of the analog memory stores the downsampled image, while the second half stores a version of the image shifted by eight columns to the left [Fig.~\ref{fig:10_mac_units_operation_principle}(a)]. This routing from the outputs of the DS3 units to the analog memory is ensured by switches changing the connections depending on the DS factor. When computing the convolution operation, this storage pattern of the image into the analog memory allows to improve throughput, by executing in parallel operations corresponding to two different shifts of the image in the execution without DS [Fig.~\ref{fig:10_mac_units_operation_principle}(c)]. The throughput is thereby increased by the DS factor, here 2$\times$. The same reasoning holds for a DS by 4$\times$ for which three shifted versions of the image are used, as shown in Fig.~\ref{fig:10_mac_units_operation_principle}(a).

\subsubsection{Switched-Cap Amplifiers for Multiplication}
We now zoom in on the convolution operation between a 16$\times$16 image patch and a 4b 16$\times$16 filter, computed by the process depicted in Fig.~\ref{fig:11_mac_units_schematic} using an SC amplifier. Phase $\raisebox{.5pt}{\textcircled{\raisebox{-.9pt} {1}}}$ of this process, presented in Fig.~\ref{fig:11_mac_units_schematic}(a), consists in successively computing the psums resulting from the convolution of a row of pixels stored in the analog memory with the corresponding row of 4b filter weights stored in the LMEM. Each psum is stored in a 16th of the SAR ADC capacitive DAC (CDAC), until all psums have been computed. In phase $\raisebox{.5pt}{\textcircled{\raisebox{-.9pt} {2}}}$, the CDAC is disconnected from the SC amplifier (\texttt{VIN\_CONNECT} = 0) and all capacitors storing psums are shorted together to compute the final convolution result by charge sharing on node $V_\mathrm{SH}$.\\
\indent Going one step further, the psum of a row is computed by the SC amplifier circuit drawn in Fig.~\ref{fig:11_mac_units_schematic}(b), whose timing diagram is detailed in Fig.~\ref{fig:11_mac_units_schematic}(c). In step $\raisebox{.5pt}{\textcircled{\raisebox{-.9pt} {1}}}$, the OTA, based on a two-stage Miller architecture, is enabled and its feedback is activated. As for the DS3 units, power gating the OTA is a key feature to save energy and improve the EE of the accelerator. Then, steps $\raisebox{.5pt}{\textcircled{\raisebox{-.9pt} {2}}}$ and $\raisebox{.5pt}{\textcircled{\raisebox{-.9pt} {3}}}$ respectively consist in resetting the columns of the analog memory corresponding to positive-weighted inputs and in reading these inputs from the analog memory. Next, in step $\raisebox{.5pt}{\textcircled{\raisebox{-.9pt} {4}}}$, the columns of the analog memory corresponding to a negative weight are reset, while connecting them to the input of the corresponding MAC units.
Lastly, during step $\raisebox{.5pt}{\textcircled{\raisebox{-.9pt} {5}}}$, the negative-weighted inputs are read, and charges are dumped on node $V_\mathrm{AMP,IN-}$ by caps $C_\mathrm{+,i}$ and $C_\mathrm{-,j}$, yielding an output voltage $V_\mathrm{MAC}$ containing the psum $\Delta V_\mathrm{CONV}$ referred to $V_\mathrm{CM}$. The formula given in Fig.~\ref{fig:11_mac_units_schematic}(b) can be intuitively understood by noticing that the inputs of caps $C_{+,i}$ are applied when $\phi_\mathrm{1,SC}$ = 1, while the inputs of caps $C_\mathrm{-,j}$ are applied when $\phi_\mathrm{2,SC}$ = 1. The inputs associated with $C_\mathrm{+,i}$ thus follow the behavior of a non-inverting SC amplifier, while those associated with $C_\mathrm{-,j}$ follow that of an inverting one, thus explaining the formula for $V_\mathrm{MAC}$. Despite the fact that the computation is performed in the mixed-signal domain and suffers from analog nonidealities, the proposed structure features several properties ensuring the robustness of the computation. (i) It has a single-ended output which does not rely on intermediate differential voltages, avoiding an incorrect result when the differential voltage is small but the common mode is large and potentially subject to saturation. This is an issue encountered in previous charge-domain near-sensor architectures \cite{Bong_2018}. (ii) The proposed structure is ratiometric and robust to PVT variations. (iii) It is not impacted by the statistical offset of the OTA thanks to the offset-insensitive switching scheme. Indeed, the charges at node $V_\mathrm{AMP,IN-}$ are
\begin{IEEEeqnarray}{L}
	Q_\mathrm{1} = -\sum_{i=1}^{N_\mathrm{+}}C_\mathrm{+,i}(V_\mathrm{BUF+,i}-V_\mathrm{AMP,IN-})\IEEEnonumber\\
	- \sum_{j=1}^{N_\mathrm{-}}C_\mathrm{-,j}(-V_\mathrm{AMP,IN-}) + C_\mathrm{FB} (V_\mathrm{AMP,IN-}-V_\mathrm{CM})
\end{IEEEeqnarray}
for $\phi_\mathrm{1,SC}$ = 1, where $N_\mathrm{+}$ and $N_\mathrm{-}$ respectively stand for the number of positive- and negative-weighted inputs, and
\begin{figure}[!t]
	\centering
	\includegraphics[width=.45\textwidth]{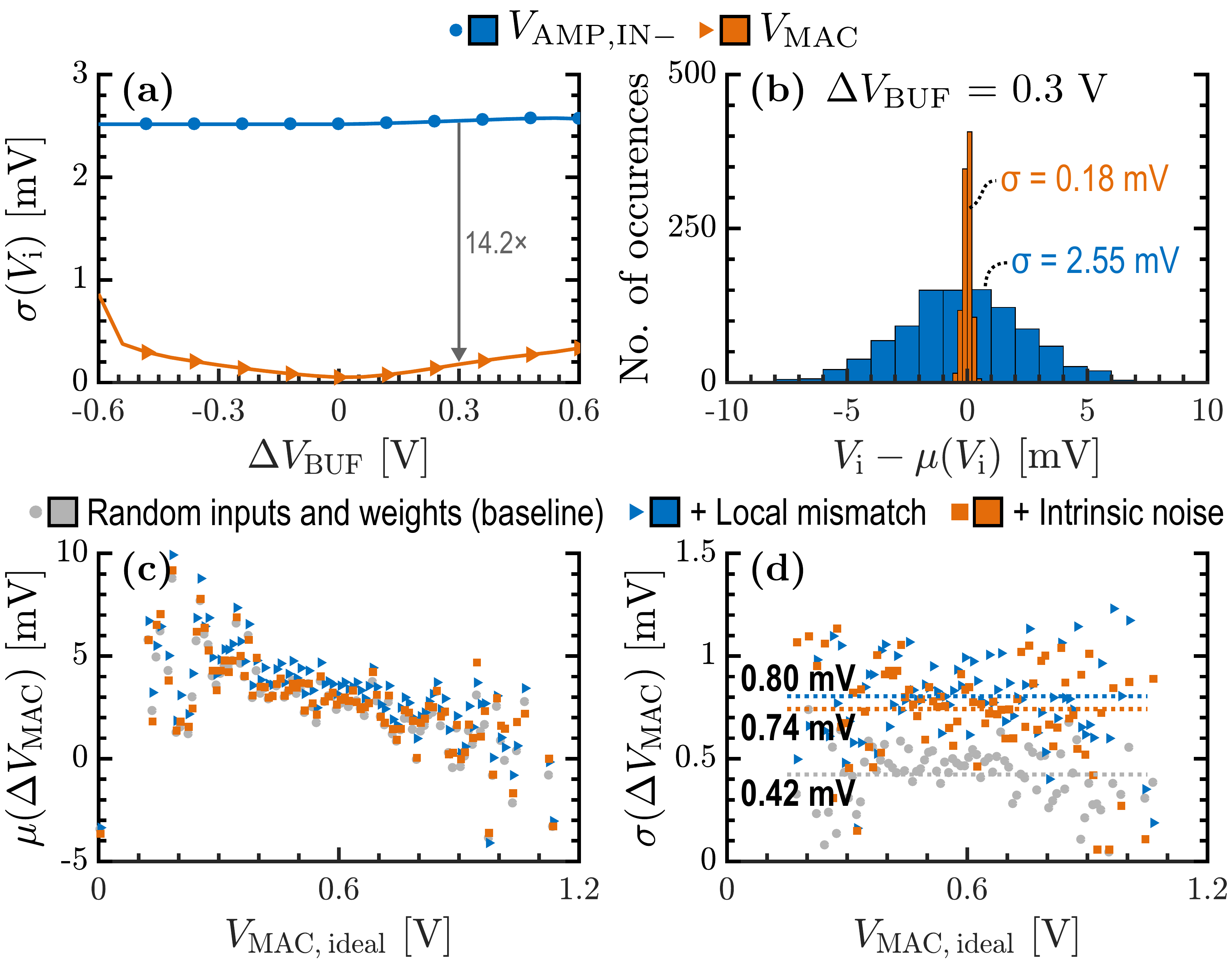}
	\caption{(a) Standard deviation and (b) distribution of $V_{\mathrm{AMP,IN-}}$ and $V_{\mathrm{OUT}}$, for 10$^3$ MC simulations with local mismatch. (b) depicts the histograms for $\Delta V_{\mathrm{BUF}}$ = 0.3~V. (c) Mean and (d) standard deviation of the error $\Delta V_{\mathrm{MAC}} = (V_{\mathrm{MAC}}-V_{\mathrm{MAC,\:ideal}})$ for 10$^4$ random combinations of inputs and weights without mismatch and noise, with only local mismatch, and with only intrinsic noise. All figures correspond to the TT 25~$^\circ$C corner.}
	\label{fig:12_mac_units_charac}
\end{figure}
\begin{figure}[!t]
	\centering
	\includegraphics[width=.5\textwidth]{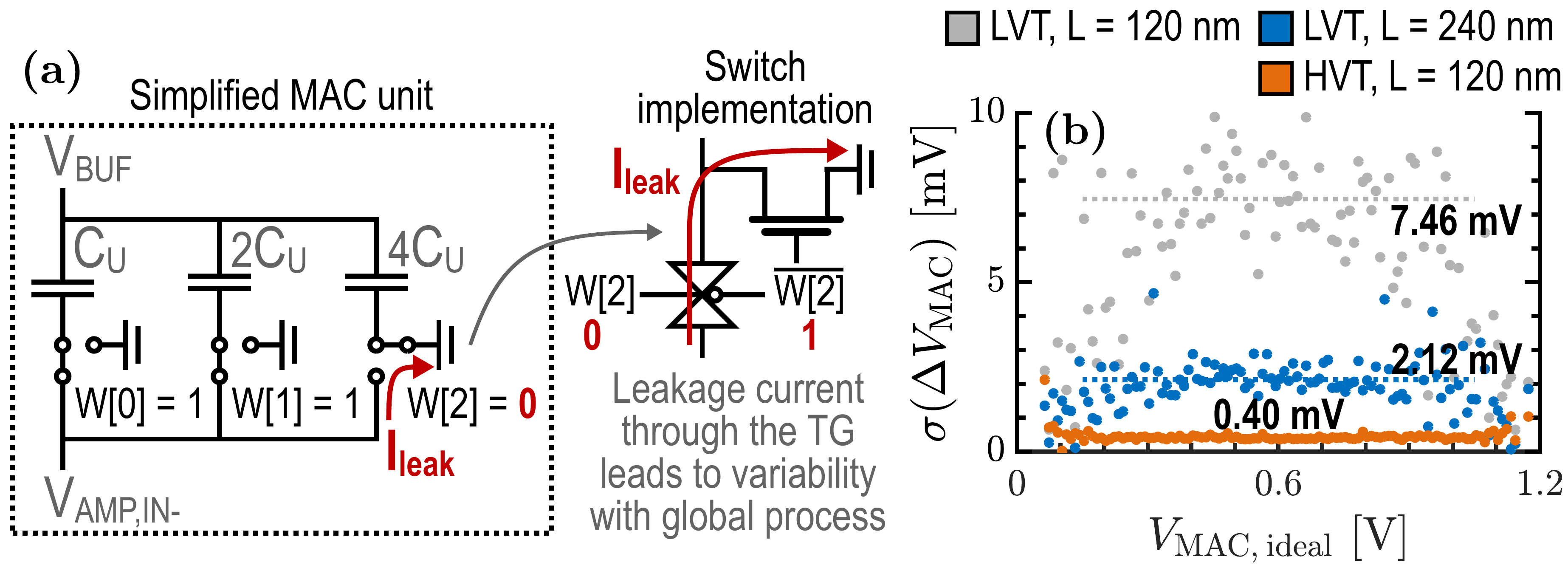}
	\caption{Leakage current through TGs in the MAC unit can lead to variability of $V_\mathrm{MAC}$ due to global process variations. (a) Simplified schematic of the MAC unit with transistor-level switch implementation, illustrating the origin of this leakage, and (b) standard deviation of the error $\Delta V_\mathrm{MAC}$ for 10$^4$ random combinations of inputs and weights with local mismatch and global process variations, for TGs realized with LVT core devices with $L$ = 120 and 240~nm, or HVT core ones with $L$ = 120~nm.}
	\label{fig:13_mac_units_improvement}
\end{figure}
\begin{figure*}[!t]
	\centering
	\includegraphics[width=\textwidth]{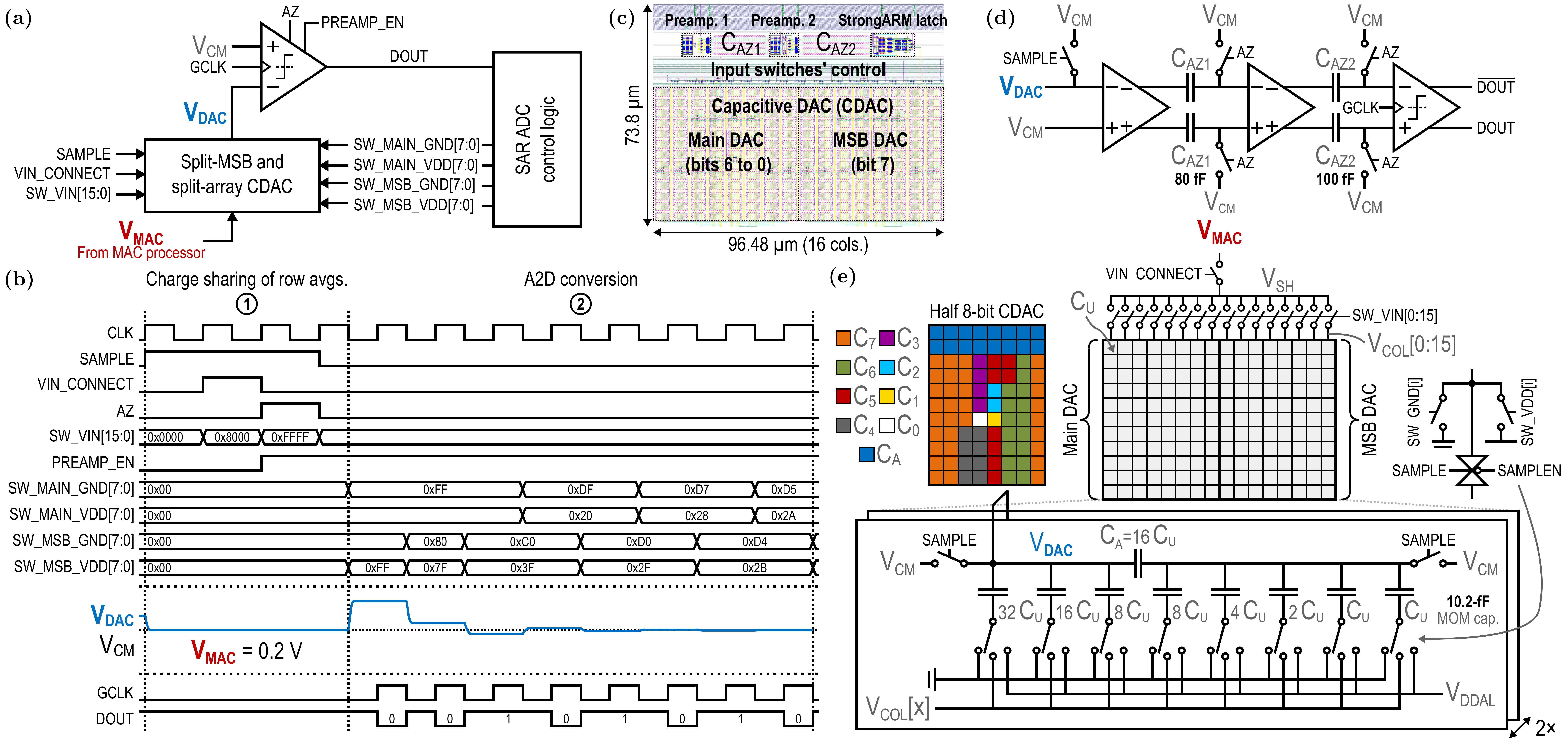}
	\caption{(a) Schematic, (b) timing diagram, and (c) layout of the 8b SAR ADC spanning over 16 columns of the analog memory. Detailed schematic of (d) the dynamic comparator, consisting of two preamplification stages and a StrongARM latch, and (e) the capacitive DAC, employing the split-MSB and split-array techniques.}
	\label{fig:14_sar_adc_schematic}
\end{figure*}
\begin{IEEEeqnarray}{L}
	Q_\mathrm{2} = -\sum_{j=1}^{N_\mathrm{-}}C_\mathrm{-,j}(V_\mathrm{BUF-,j}-V_\mathrm{AMP,IN-})\IEEEnonumber\\
	 - \sum_{i=1}^{N_\mathrm{+}}C_\mathrm{+,i}(-V_\mathrm{AMP,IN-}) + C_\mathrm{FB} (V_\mathrm{AMP,IN-}-V_\mathrm{MAC})
\end{IEEEeqnarray}
for $\phi_\mathrm{2,SC}$ = 1. Interestingly, the resulting expression for $V_\mathrm{MAC}$ based on the conservation of charge at node $V_\mathrm{AMP,IN-}$, i.e., $Q_\mathrm{1} = Q_\mathrm{2}$, does not depend on the value of $V_\mathrm{AMP,IN-}$ when $\phi_\mathrm{1,SC}$ = 1 and therefore, is independent of the OTA's offset.\\
\indent Furthermore, the implementation of the 4b weights is given in Fig.~\ref{fig:11_mac_units_schematic}(d), with the most-significant bit (MSB) \texttt{W[3]} corresponding to the sign bit and least-significant bits (LSBs) \texttt{W[2:0]} to the magnitude bits. The sign bit determines which signals control the connections at the input of the MAC unit, while the magnitude bits determine the number of \mbox{7-fF} unitary MOM caps $C_\mathrm{U}$ connected in parallel in each MAC unit. This circuit thus implements integer weights ranging from -7 to 7, multiplied by a factor 0.25$\times$, originating from the fact that each column includes a part of the feedback cap $C_\mathrm{FB}$ equal to 4$C_\mathrm{U}$. The MAC unit fits inside the pixel pitch and occupies a height of 28.85~$\mu$m [Fig.~\ref{fig:11_mac_units_schematic}(e)].\\
\indent The performance of these MAC units is characterized with post-layout simulations in Fig.~\ref{fig:12_mac_units_charac}. First, Figs.~\ref{fig:12_mac_units_charac}(a) and (b) correspond to a setup in which eight MAC units have their weight set to +7 with a shared input voltage $V_\mathrm{BUF,+}$, while the other eight units have their weight set to -7 with input voltage $V_\mathrm{BUF,-}$. The impact of local mismatch is studied in this context with 10$^3$ Monte Carlo (MC) simulations. Across the input range $\Delta V_\mathrm{BUF} = (V_\mathrm{BUF,+}-V_\mathrm{BUF,-})$, the $\sigma$ of $V_\mathrm{MAC}$ remains below 1~mV while that of $V_\mathrm{AMP,IN-}$ is affected by the statistical offset of the OTA, leading to a \mbox{2.5-mV} $\sigma$. More specifically, for $\Delta V_\mathrm{BUF}$ = 0.3~V, the proposed offset-insensitive structure reduces the $\sigma$ by 14.2$\times$ from 2.55 to 0.18~mV.
However, this first setup only accurately describes a specific realization of input voltages and weights. Therefore, Figs.~\ref{fig:12_mac_units_charac}(c) and (d) extend this analysis to a baseline of 10$^4$ random combinations of inputs and weights drawn from uniform distributions, with local mismatch and intrinsic noise subsequently applied on top of it. In the $V_\mathrm{MAC}$ output range between 0.15 and 1.05~V, avoiding transistors to be biased outside of saturation, the average error in Fig.~\ref{fig:12_mac_units_charac}(c) shows a deterministic behavior hinting at a slope error which can be related to parasitic capacitances or charge injection, while the average standard deviation of the error over the considered output range increases from 0.42~mV to respectively 0.80 and 0.74~mV with local mismatch and intrinsic noise. These variabilities are predominantly due to the mismatch of MOM caps and the thermal $kT/C$ noise due to the sampling of input voltages on the capacitors in the MAC units. These results demonstrate the robustness of the proposed architecture to these analog nonidealities. However, Fig.~\ref{fig:13_mac_units_improvement} highlights one of the architectural details which could be further improved. As emphasized in Fig.~\ref{fig:13_mac_units_improvement}(a), when one of the magnitude bits is equal to zero, a leakage current flows through the TG, implemented with high-speed low-$V_\mathrm{th}$ (LVT) core devices with a minimum length of 120~nm. Hence, this introduces a stochastic error on $V_\mathrm{MAC}$ when both local mismatch and global process variations are considered, with an average $\sigma$ around 7.46~mV [Fig.~\ref{fig:13_mac_units_improvement}(b)] compared to 0.80~mV with mismatch only. To mitigate this problem, the transistors constituting the TG can either be made longer or rely on low-leakage high-$V_\mathrm{th}$ (HVT) core devices, respectively reducing the average $\sigma$ to 2.12 and 0.40~mV. Interestingly, the \mbox{0.40-mV} $\sigma$ obtained with an HVT TG is even lower than the one obtained with local mismatch only with the LVT TG, suggesting that the sensitivity to local mismatch not only stems from MOM caps, but also from the TG leakage.

\subsubsection{SAR ADCs for Aggregation and Digitization}
The SAR ADCs employed in this work follow the topology presented in Fig.~\ref{fig:14_sar_adc_schematic}(a) with the associated timing diagram shown in Fig.~\ref{fig:14_sar_adc_schematic}(b). Similarly to the SC amplifiers, each SAR ADC spans 16 pixel columns [Fig.~\ref{fig:14_sar_adc_schematic}(c)]. The two main functions of the SAR ADC employed in this work are $\raisebox{.5pt}{\textcircled{\raisebox{-.9pt} {1}}}$ the aggregation of psums by charge sharing, following the calculation of the psums of rows by the SC amplifiers, and $\raisebox{.5pt}{\textcircled{\raisebox{-.9pt} {2}}}$ the analog-to-digital (A2D) conversion of the convolution result, following the SAR principle. In terms of circuit implementation, the detailed comparator architecture in Fig.~\ref{fig:14_sar_adc_schematic}(d) features two differential preamplification stages based on a differential pair driving a load of diode-connected transistors, and providing a total gain of approximately 10~V/V. These stages both embed AZ capabilities based on caps at their output, and are followed by a dynamic StrongARM latch. In addition, the CDAC combines two existing techniques to reduce its power and area overheads. First, it relies on a split-MSB array \cite{Ginsburg_2005}, based on two identical DACs for the MSB and all the LSBs, respectively called MSB DAC and main DAC in Figs.~\ref{fig:14_sar_adc_schematic}(c) and (e). More specifically, as illustrated in Fig.~\ref{fig:14_sar_adc_schematic}(b), the MSB DAC is initially switched to $V_\mathrm{DD}$. Then, when the output data \texttt{DOUT} of the comparator is equal to zero, the current bit of the MSB DAC is switched to ground, whereas when \texttt{DOUT} is equal to one, the current bit of the main DAC is switched to $V_\mathrm{DD}$.
\begin{figure}[!t]
	\centering
	\includegraphics[width=.45\textwidth]{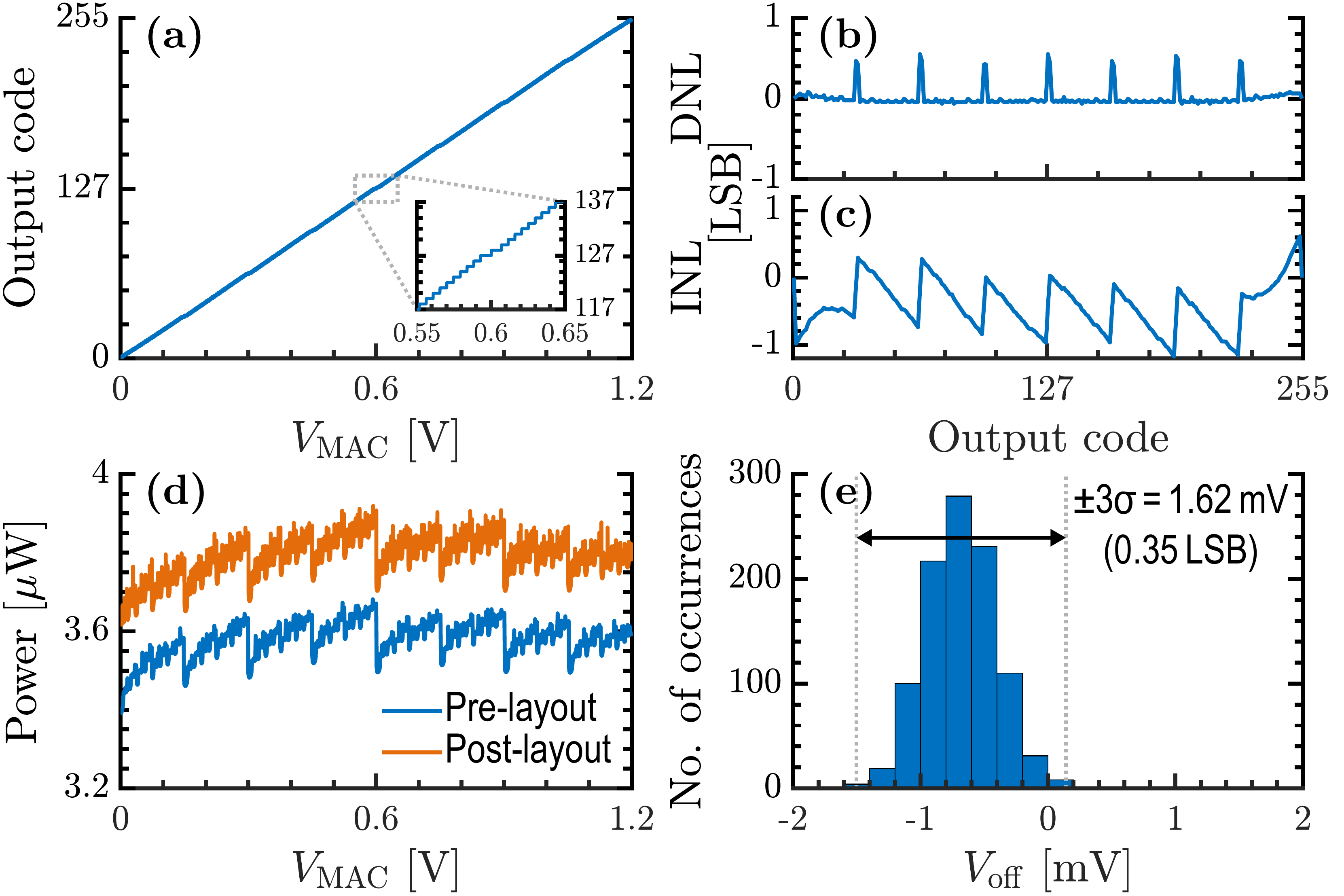}
	\caption{(a) Transfer function of the SAR ADC with the corresponding (b) DNL and (c) INL. (d) Power consumption on $V_{\mathrm{DDAL}}$ = 1.2~V (CDAC, comparator, and drivers) as a function of the input voltage. (e) Statistical offset of the StrongARM latch referred to the input of the first preamplifier. All figures correspond to the TT 25~$^\circ$C corner.}
	\label{fig:15_sar_adc_charac}
\end{figure}
This allows to harmonize power consumption across the input voltage range by optimizing the switching of the DAC. Then, it also makes use of a split-capacitive array \cite{Li_2014}, employing an attenuation cap $C_\mathrm{A}$ in Fig.~\ref{fig:14_sar_adc_schematic}(e) to reduce the impact of the LSBs, ultimately allowing to diminish the capacitance of the MSBs.  Besides, possible further improvements in terms of EE and silicon area can be obtained by employing parasitic caps instead of explicit MOM ones, as outlined by Harpe in \cite{Harpe_2019}. Interestingly, in the RoI detection mode generating 1b fmaps, the offset associated with each filter is also implemented with the CDAC, by switching up (resp. down) bits of the main (resp. MSB) DAC to implement a positive (resp. negative) offset.\\
\indent Regarding the post-layout simulation results in typical conditions (TT 25~$^\circ$C) in Fig.~\ref{fig:15_sar_adc_charac}, a relatively good linearity is achieved in Fig.~\ref{fig:15_sar_adc_charac}(a), with the differential nonlinearity (DNL) comprised between -0.07 and 0.55~LSB, and the integral one (INL) between -1.17 and 0.62~LSB [Figs.~\ref{fig:15_sar_adc_charac}(b) and (c)]. Regarding power consumption, it is relatively constant across the input voltage range thanks to the split-MSB CDAC, with a mean value of 3.59 and 3.78~$\mu$W in pre- and post-layout simulations [Fig.~\ref{fig:15_sar_adc_charac}(d)]. At last, the input-referred statistical offset of the comparator features a $\pm$3$\sigma$ value of 1.62~mV corresponding to 0.35~LSB, hence ensuring that the comparator operation is robust to local mismatch [Fig.~\ref{fig:15_sar_adc_charac}(e)].

\section{Experimental Results}
\label{sec:4_experimental_results}
\begin{figure}[!t]
	\centering
	\includegraphics[width=.5\textwidth]{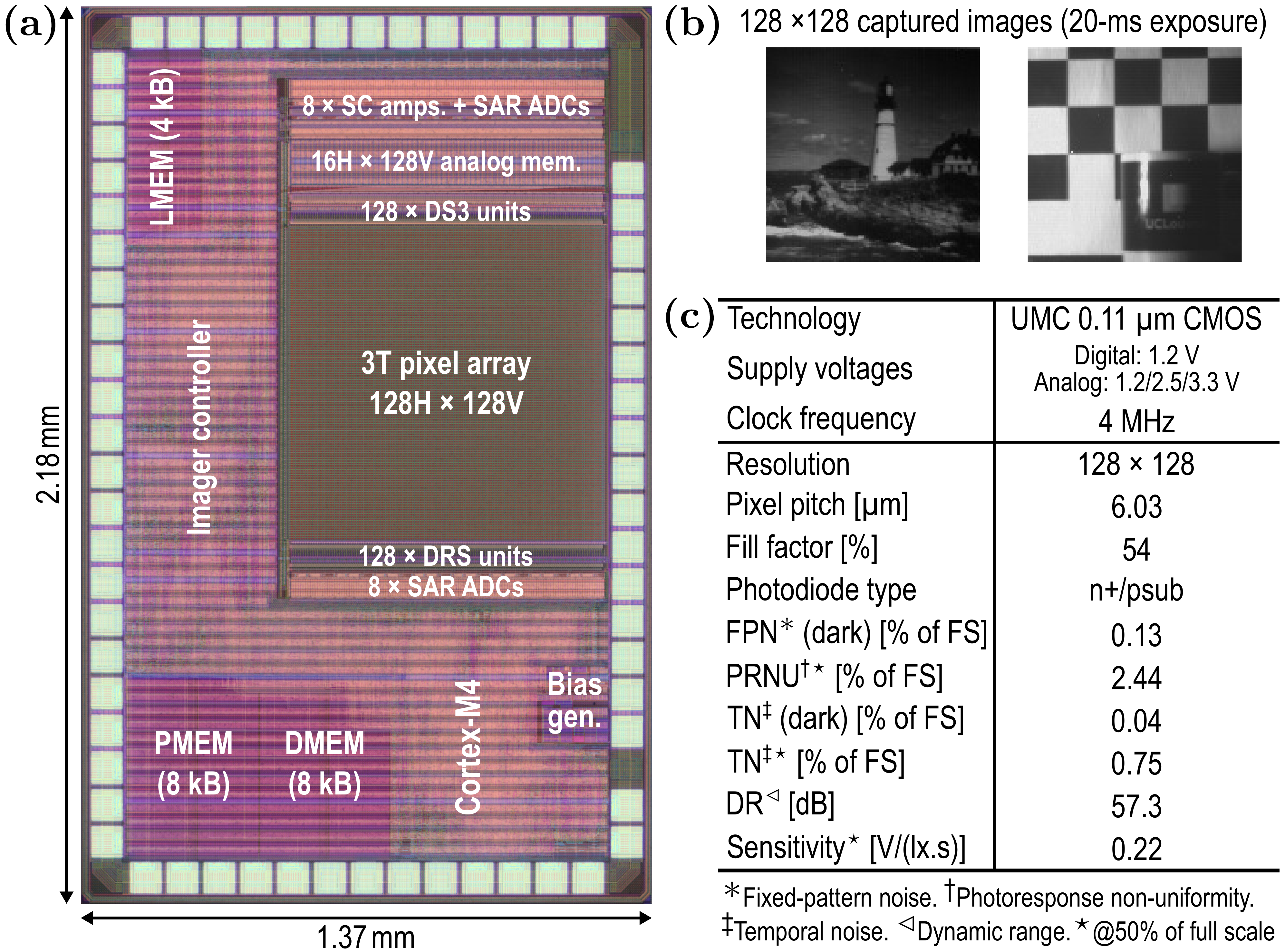}
	\caption{(a) Chip microphotograph with overlaid layout. (b) Example of captured images with an exposure time of 20~ms and (c) imaging characteristics of the proposed imager.}
	\label{fig:16_chip_microphotograph}
\end{figure}
\begin{figure}[!t]
	\centering
	\includegraphics[width=.45\textwidth]{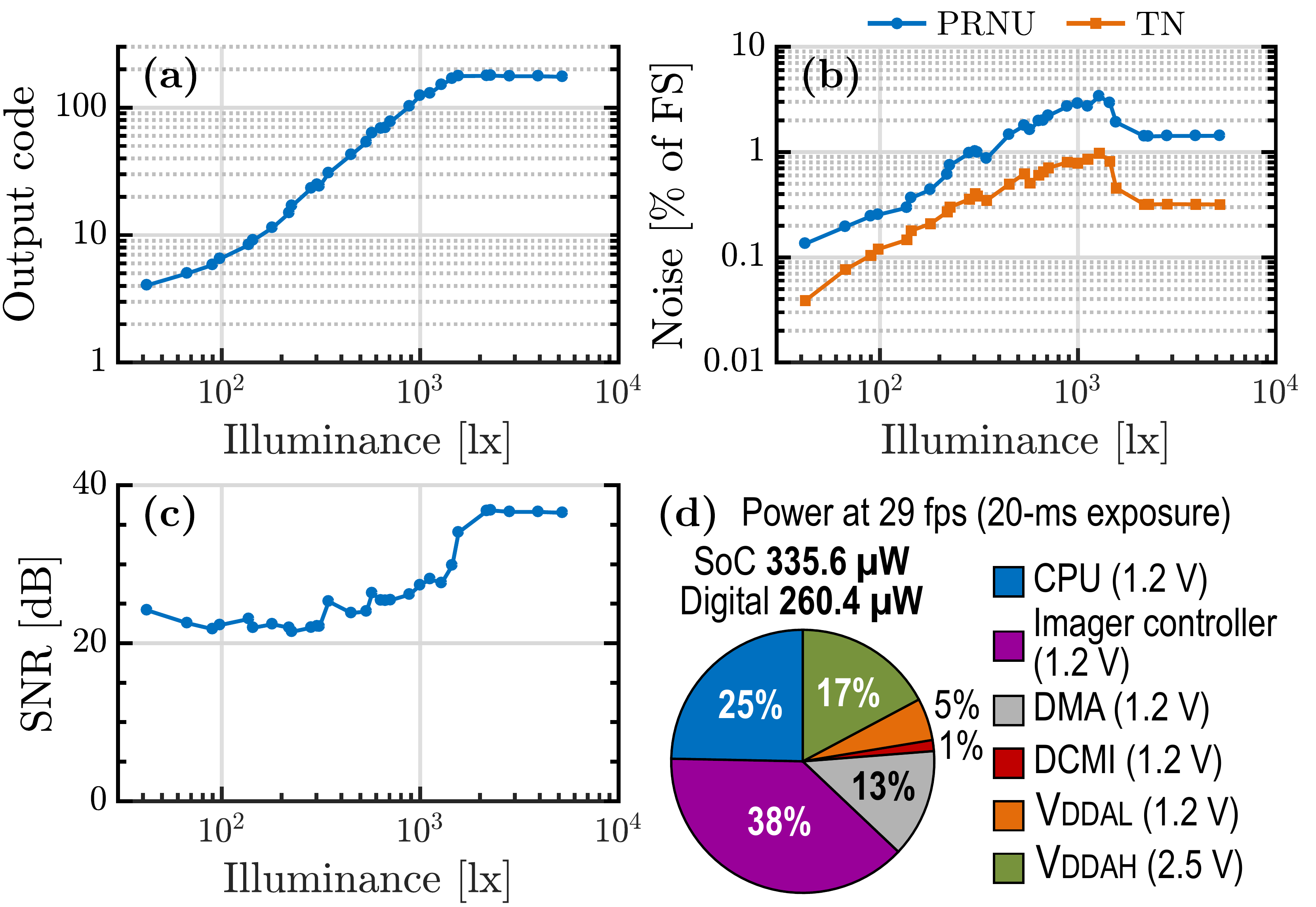}
	\caption{(a) Output code, (b) PRNU and TN, and (c) SNR as a function of illuminance, computed for 10 images for each illuminance. (d) Power breakdown in imaging mode. All figures correspond to an exposure of 20~ms.}
	\label{fig:17_meas_image_mode}
\end{figure}
MANTIS imager SoC has been fabricated in UMC \mbox{0.11-$\mu$m} bulk CMOS technology. The chip microphotograph is shown in Fig.~\ref{fig:16_chip_microphotograph}(a), together with examples of captured images in Fig.~\ref{fig:16_chip_microphotograph}(b). This section presents the experimental results obtained with the fabricated chip, respectively focusing on the characterization of the image sensor in Section~\ref{subsec:4A_imaging_performance} and of the mixed-signal near-sensor convolution processor in Section~\ref{subsec:4B_electrical_characterization}. Finally, the applicative performance is evaluated on a face RoI detection task in Section~\ref{subsec:4C_face_RoI_detection}.
\begin{figure*}[!t]
	\centering
	\includegraphics[width=\textwidth]{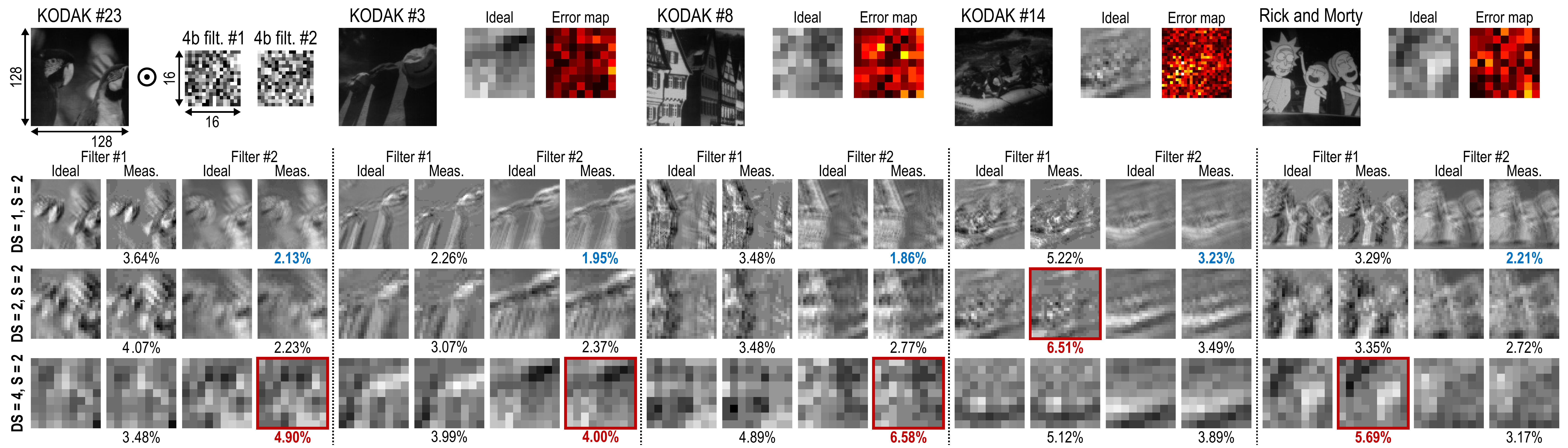}
	\caption{Comparison between ideal and measured fmaps, respectively computed in software (Matlab) and on chip, obtained by a convolution operation between the image and two random 4b 16$\times$16 filters. The parameters used for this operation are S = 2, DS = 1, 2 or 4, and an exposure time of 12.5~ms. The RMSE for each fmap is written below it, and the error map corresponds to the fmap with the worst RMSE among the displayed ones.}
	\label{fig:18_meas_conv_mode_fmaps}
\end{figure*}

\subsection{Imaging Performance}
\label{subsec:4A_imaging_performance}
We first characterize the performance of MANTIS in imaging mode with the results summarized in Figs.~\ref{fig:16_chip_microphotograph}(c) and \ref{fig:17_meas_image_mode}. This characterization is performed by exposing the imager to an uniform light flux, generated by an Olympus KL 1500 halogen light source going through an integrating sphere, and by capturing 10 images for each light flux level. The transfer function between the imager 8b output code and the light flux per unit area, i.e., the illuminance expressed in lm/m$^2$ or lx, is depicted in Fig.~\ref{fig:17_meas_image_mode}(a). It highlights that, for a \mbox{20-ms} exposure time, the usable illuminance ranges from 120 to 1500~lx. In addition, the transfer function levels off at low illuminance due to the relatively low photocurrent values compared to leakages inside the pixel, the photoresponse of the n+/psub diodes available in this CMOS logic process being far from optimal compared to a CMOS image sensor (CIS) process. Next, Fig.~\ref{fig:17_meas_image_mode}(b) evaluates two types of noise affecting the image quality, namely photoresponse non-uniformity (PRNU), capturing the variability of pixel responses to light due to mismatch, and temporal noise (TN). PRNU and TN are worth 2.44 and 0.75$\%$ of the full scale (FS) at 50$\%$ of the FS, and correspond to a signal-to-noise ratio (SNR) slightly above 20~dB in the usable illuminance range, dominated by the PRNU [Fig.~\ref{fig:17_meas_image_mode}(c)]. Taking a more theoretical perspective, the voltage noise due to thermal noise at the output of the DRS units can be expressed as
\setlength{\tabcolsep}{2pt}
\begin{table*}[!t]
\centering
\caption{Summary of the measured performance of MANTIS imager SoC in convolution mode for four filters (\mbox{12.5-ms} exposure time).}
\label{table:summary_measured_performance}
\vspace{-0.15cm}
%\scalebox{.8}{%
\begin{threeparttable}
\begin{scriptsize}
\begin{tabular}{@{}p{\textwidth}@{}}
\centering
\begin{tabular}[t]{l|cccc|cccc|cccc|}
\toprule
Image downsampling (DS) & \multicolumn{4}{c}{1} & \multicolumn{4}{c}{2} & \multicolumn{4}{c}{4}\\
\cmidrule(lr){2-5} \cmidrule(lr){6-9} \cmidrule(lr){10-13}
Filter stride (S) & 2 & 4 & 8 & 16 & 2 & 4 & 8 & 16 & 2 & 4 & 8 & 16\\
\midrule
Frame rate\tnote{$\star$}\quad[fps] & 18.2 & 79.7 & 79.7 & 79.7 & 79.7 & 79.7 & 79.7 & 79.7 & 79.7 & 79.7 & 79.7 & 79.7\\
Throughput\tnote{$\dagger$}\quad[MOPS] & 121 & 137.3 & 36.7 & \textcolor{ECS-Red}{\textbf{10.5}} & \textcolor{ECS-Blue}{\textbf{408.3}} & 110.4 & 32.0 & \textcolor{ECS-Red}{\textbf{10.5}} & 211.7 & 65.3 & 23.5 & \textcolor{ECS-Red}{\textbf{10.5}}\\
Feature map RMSE\tnote{$\diamond$}\quad[$\%$] & \textcolor{ECS-Blue}{\textbf{3.01}} & 3.25 & 4.00 & 4.69 & 3.40 & 3.98 & 6.30 & 8.68 & 4.88 & \textcolor{ECS-Red}{\textbf{11.34}} & 9.19 & 8.45\\
\midrule
Power\tnote{$\triangleright$}\quad(accelerator) [$\mu$W] & 66.84 & \textcolor{ECS-Red}{\textbf{76.20}} & 22.36 & 8.40 & 58.74 & 17.40 & 6.60 & 4.03 & 10.07 & 4.42 & 3.29 & \textcolor{ECS-Blue}{\textbf{2.70}}\\
EE\tnote{$\triangleleft$}\quad(accelerator) [TOPS/W] & 7.24 & 7.31 & 6.57 & \textcolor{ECS-Red}{\textbf{4.98}} & 27.80 & 25.38 & 19.40 & 10.37 & \textcolor{ECS-Blue}{\textbf{84.09}} & 59.17 & 28.61 & 15.48\\
Energy/OP\tnote{$\triangleleft$}\quad(accelerator) [fJ/op] & 138.1 & 138.7 & 152.1 & \textcolor{ECS-Red}{\textbf{200.9}} & 36.0 & 39.4 & 51.6 & 96.4 & \textcolor{ECS-Blue}{\textbf{11.9}} & 16.9 & 35.0 & 64.6\\
\midrule
Power\tnote{$\ddagger$}\quad(SoC) [$\mu$W] & 338.5 & \textcolor{ECS-Red}{\textbf{384.7}} & 297.4 & 268.9 & 357.0 & 288.0 & 264.7 & 256.3 & 271.9 & 258.3 & 253.3 & \textcolor{ECS-Blue}{\textbf{250.9}}\\
EE\tnote{$\triangleleft$}\quad(SoC) [TOPS/W] & 1.43 & 1.43 & 0.49 & \textcolor{ECS-Red}{\textbf{0.16}} & \textcolor{ECS-Blue}{\textbf{4.57}} & 1.53 & 0.48 & \textcolor{ECS-Red}{\textbf{0.16}} & 3.11 & 1.01 & 0.37 & 0.17\\
Energy/OP\tnote{$\triangleleft$}\quad(SoC) [pJ/op] & 0.70 & 0.70 & 2.02 & \textcolor{ECS-Red}{\textbf{6.43}} & \textcolor{ECS-Blue}{\textbf{0.22}} & 0.65 & 2.07 & \textcolor{ECS-Red}{\textbf{6.13}} & 0.32 & 0.99 & 2.69 & 6.00\\
Processing energy (SoC) [pJ/(pix$\cdot$frame$\cdot$filt)] & \textcolor{ECS-Red}{\textbf{284.1}} & 73.6 & 56.9 & 51.5 & 68.3 & 55.1 & 50.7 & 49.0 & 52.0 & 49.4 & 48.5 & \textcolor{ECS-Blue}{\textbf{48.0}}\\
\bottomrule
\end{tabular}
\end{tabular}
\end{scriptsize}
\begin{footnotesize}
\begin{tablenotes}
	\item[$\star$] Frame rate is limited by the \mbox{12.5-ms} exposure time.
	\item[$\dagger$] Expressed in operations with analog inputs and 4b weights.
	\item[$\diamond$] Computed over 10 images with 10 random filters.
	\item[$\triangleright$] Includes the analog memory, SC amplifiers, SAR ADCs and drivers on $V_\mathrm{DDAL}$.
	\item[$\triangleleft$] Normalized to 1b operations.
	\item[$\ddagger$] Includes the imager analog macro ($V_\mathrm{DDAL}$ and $V_\mathrm{DDAH}$), and the digital core, i.e., the Cortex-M4 CPU, the imager controller, and the SRAM macros ($V_\mathrm{DDD}$).
\end{tablenotes}
\end{footnotesize}
\end{threeparttable}%
%}
\end{table*}
\begin{equation}
	\overline{v_\mathrm{n}} = \sqrt{2kT} \sqrt{\frac{A_\mathrm{SF}^2}{C_\mathrm{PD}} + \frac{1}{C_\mathrm{S}}}
\end{equation}
with $C_\mathrm{PD}$ = 12.2~fF the pixel capacitance, $C_\mathrm{S}$ = 29~fF the sampling capacitance, and $A_\mathrm{SF}$ = 0.69~V/V the gain of the pixel's SF. This yields $\overline{v_\mathrm{n}}$ = 0.78~mV at 25~$^\circ$C corresponding to a 0.65$\%$ error with a \mbox{1.2-V} dynamic range of $V_\mathrm{PIX}$, which is in the same order of magnitude as the TN in Fig.~\ref{fig:17_meas_image_mode}(b). A 4T pixel array could easily be integrated to the proposed design by switching to pinned photodiodes and modifying the digital controller, and would reduce the contribution of TN thanks to correlated double sampling (CDS). The imaging characteristics are summarized in Fig.~\ref{fig:16_chip_microphotograph}(c). Lastly, Fig.~\ref{fig:17_meas_image_mode}(d) describes the power consumption of the imager for a frame rate of 29~fps. Power is dominated by the digital part, which represents 78$\%$ of the \mbox{335.6-$\mu$W} SoC power, with the following split: 38$\%$ for the imager controller, 25$\%$ for the CPU, and 13$\%$ for data transfers by the DMA. It could easily be reduced by moderately scaling $V_\mathrm{DDD}$ to 1~V or by making use of power-gating techniques. The power of the analog circuitry only amounts to 22$\%$ of the SoC power, with most of it (17$\%$) being consumed by the pixel array and DRS units supplied at 2.5~V, while the remaining 5$\%$ correspond to SAR ADCs supplied at 1.2~V. The SoC power corresponds to an energy per pixel of 706.3~pJ/(pix$\cdot$frame), which is larger than state-of-the-art values for low-power imagers, typically ranging from 100 to 300~pJ/(pix$\cdot$frame) \cite{Park_2020}. Nonetheless, this is perfectly normal as our SoC is not optimized for imaging, and as imagers usually do not include a CPU and a DMA.\looseness=-1

\subsection{Electrical Characterization of Near-Sensor Convolutions}
\label{subsec:4B_electrical_characterization}
\begin{figure}[!t]
	\centering
	\includegraphics[width=.45\textwidth]{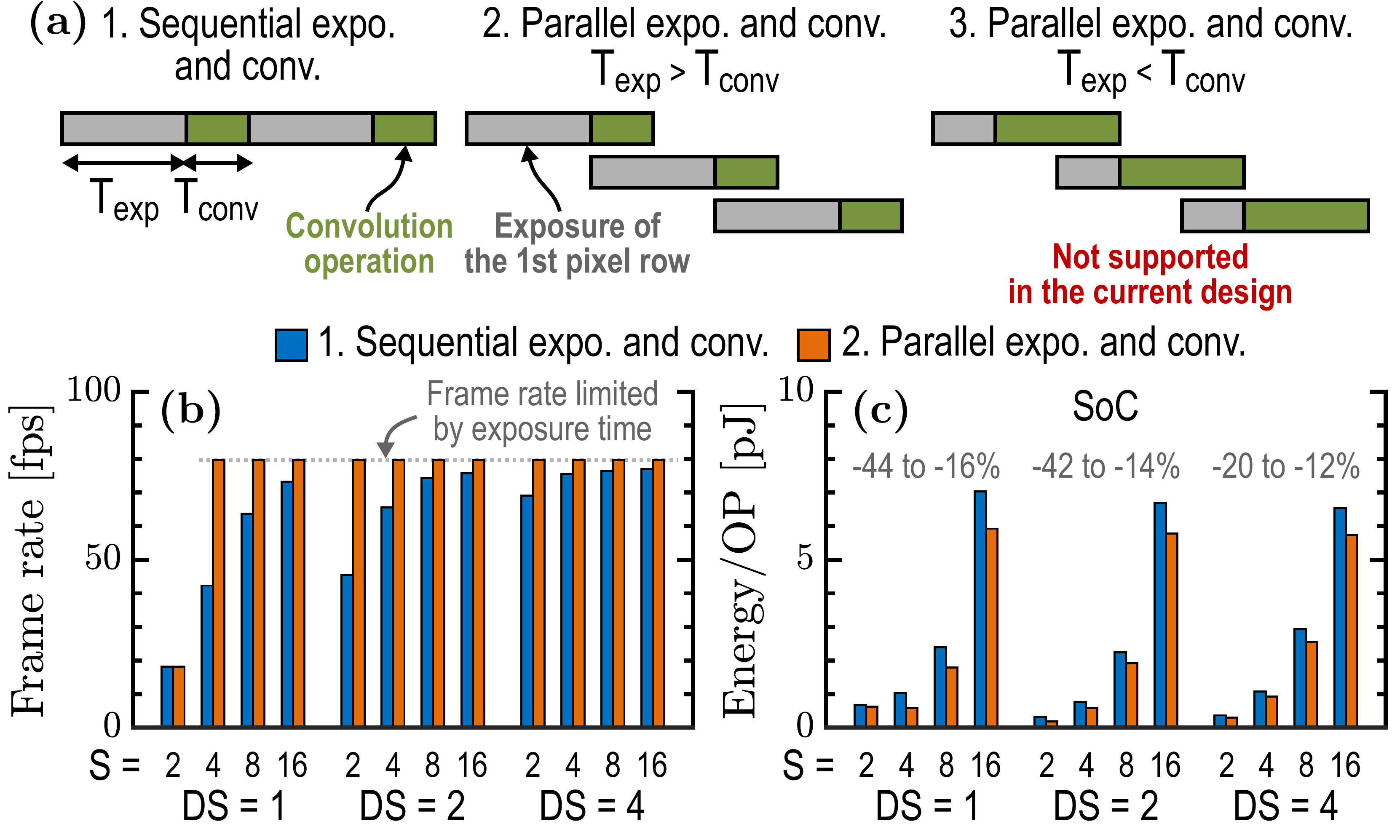}
	\caption{(a) Illustration of sequential and parallel exposure and convolution. Measured (b) frame rate and (c) energy per 1b operation (SoC) for the sequential and parallel executions and for different DS and S configurations, four filters, and a 12.5-ms exposure time.}
	\label{fig:19_meas_conv_mode_exposure}
\end{figure}
\begin{figure}[!t]
	\centering
	\includegraphics[width=.45\textwidth]{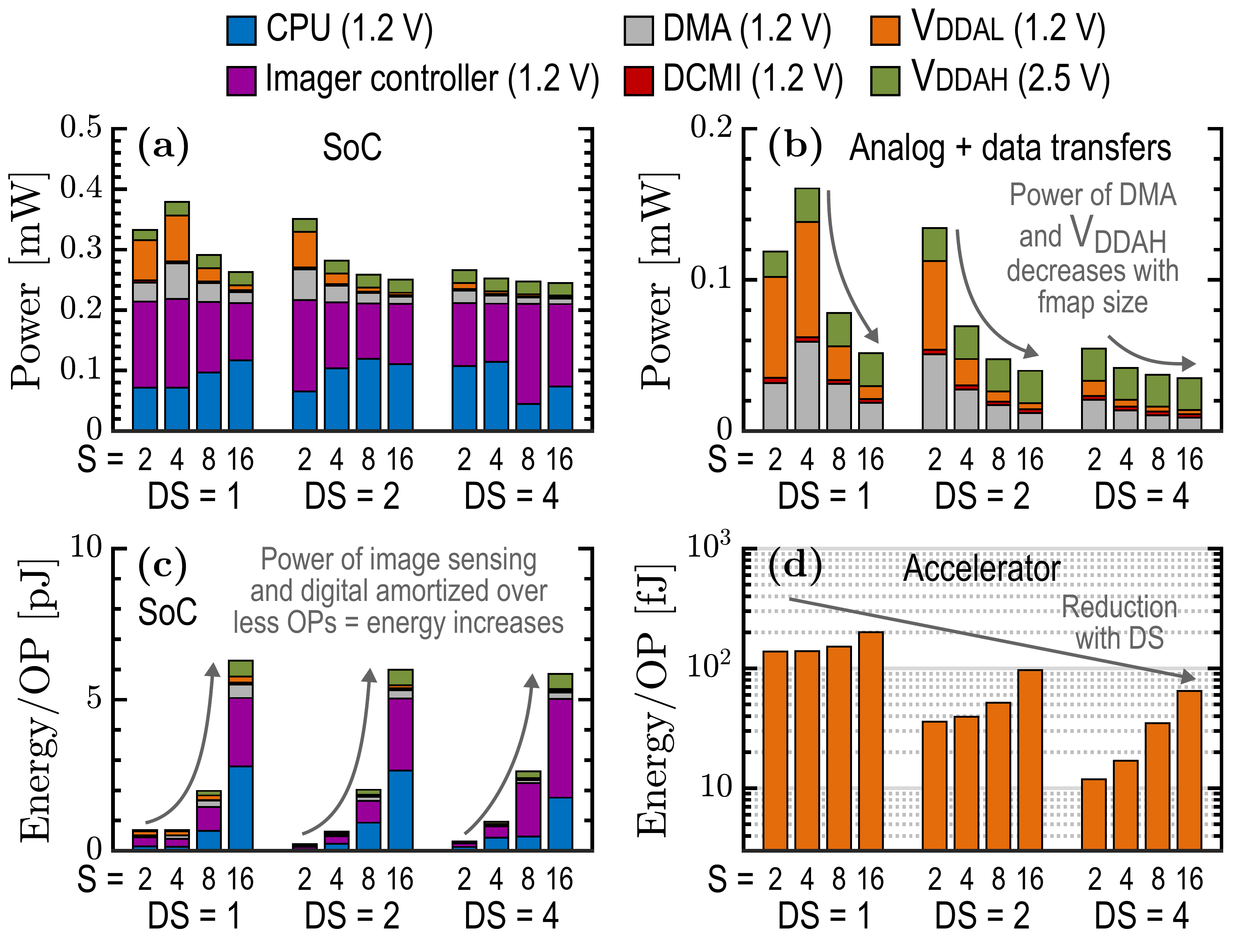}
	\caption{Breakdown of (a) the total SoC power and (b) the power of the analog macro and digital data transfers. Energy per 1b operation for (c) the SoC and (d) the accelerator. All figures correspond to a \textbf{parallel} exposure and convolution, different DS and S configurations, four filters, and a 12.5-ms exposure time, and present measurement results. The fraction of power due to the imager controller in (a)(c) is estimated based on physical simulations.}
	\label{fig:20_meas_conv_mode_power}
\end{figure}
\begin{figure}[!t]
	\centering
	\includegraphics[width=.45\textwidth]{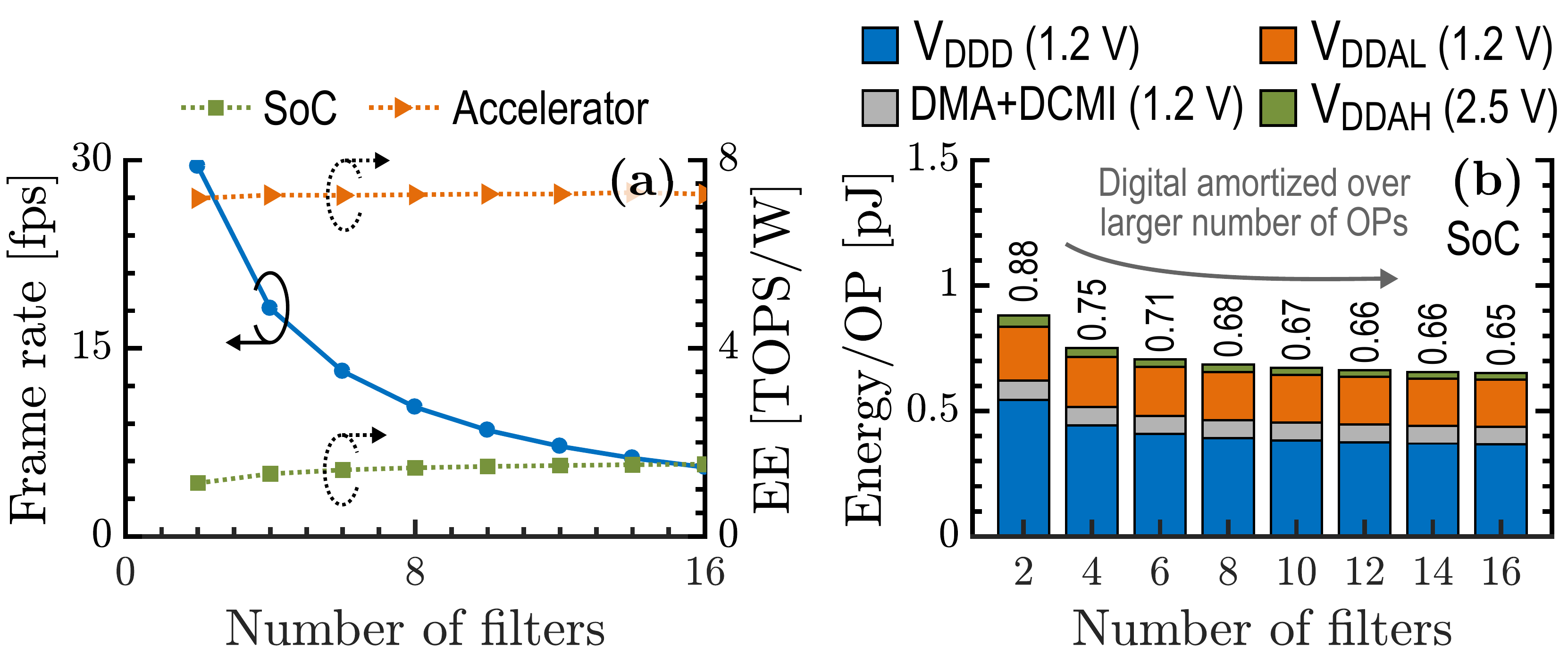}
	\caption{For \textbf{sequential} exposure and convolution, DS = 1, S = 2, and a 12.5-ms exposure time, measured (a) frame rate, EE at the SoC and accelerator levels, and (b) energy per 1b operation with the number of filters.}
	\label{fig:21_meas_conv_mode_nfilts}
\end{figure}

To validate the proper operation of the mixed-signal convolution processor, two aspects need to be thoroughly quantified: (i) the quality of the fmaps computed by the chip with respect to an ideal execution in software [Fig.~\ref{fig:18_meas_conv_mode_fmaps}], and (ii) the throughput and EE of the MAC operations, at the accelerator and SoC levels [Figs.~\ref{fig:19_meas_conv_mode_exposure} to \ref{fig:21_meas_conv_mode_nfilts}]. This analysis must cover the different configurations of the proposed convolution processor in terms of image DS factor and filter stride S. For the characterization of throughput and EE, we rely on the benchmarking outlined by Shanbhag et al. \cite{Shanbhag_2022} in the context of IMC, which shares striking similarities with convolutional imagers, except that IMC is weight-stationary while convolutional imagers are input-stationary. The concept of \textit{ADC column} originates from this paper, and corresponds in this work to a group of 16 columns of the analog memory feeding 16 MAC units, connected to an SC amplifier and an 8b SAR ADC.\\
\indent We start by comparing ideal fmaps to measured ones based on two important steps. First, fmaps need to be normalized, as the pixel values of the 8b 128$\times$128 image used in the ideal software execution in Matlab do not reflect the analog values employed on chip by the convolution processor. This difference stems from the various multiplicative factors applied to raw pixel voltages in the blocks constituting the convolution pipeline. For a given fmap denoted as $f$, the normalized fmap denoted as $\hat{f}$ is computed as
\begin{equation}
	\hat{f} = \frac{\left[f - \mu(f)\right]}{\sigma(f)}\textrm{,}\label{eq:fmap_norm}
\end{equation}
thereby ensuring that the mean $\mu$ and the standard deviation $\sigma$ of the resulting fmap are respectively equal to zero and one. The second important step is the metric used to assess the quality of the computed fmap. Here, we rely on the root mean square error (RMSE) calculated as
\begin{equation}
	\mathrm{RMSE} = \frac{\mathrm{100}\%}{2\mathrm{max}(|\hat{f}_\mathrm{meas}|)} \sqrt{\frac{1}{N_\mathrm{f}^2} \sum_{i=1}^{N_\mathrm{f}} \sum_{j=1}^{N_\mathrm{f}} \left(\hat{f}_\mathrm{ideal,ij} - \hat{f}_\mathrm{meas,ij}\right)^2}\label{eq:RMSE}
\end{equation}
where $\hat{f}_\mathrm{ideal}$ and $\hat{f}_\mathrm{meas}$ are respectively the normalized ideal and measured fmaps, and $N_\mathrm{f}$ is the fmap size, obtained from
\begin{equation}
	N_\mathrm{f} = \left(\frac{128}{\textrm{DS}}-\textrm{F}\right)\frac{1}{\textrm{S}} + 1\label{eq:fmap_size}
\end{equation}
with F = 16 the filter size. The characterization is performed over 10 images, among which nine are part of the KODAK dataset of natural images, and with 10 4b-weighted filters drawn from a uniform distribution. Table~\ref{table:summary_measured_performance} details the RMSE results. It indicates that the RMSE is comprised between 3.01 and 11.34$\%$, and that it tends to degrade for smaller fmaps with a larger DS factor and/or a larger filter stride. This is quite intuitive to understand, given that (\ref{eq:RMSE}) relies on $\mathrm{max}(|\hat{f}_\mathrm{meas}|)$ to approximate the range of values contained in an fmap, and is hence sensitive to errors in large values of $\hat{f}_\mathrm{meas}$. However, we believe that the proposed metric provides both an intuition and a quantification of the magnitude of the error, despite these inaccuracies. A few fmaps are displayed in Fig.~\ref{fig:18_meas_conv_mode_fmaps} with the corresponding RMSE, as well as an error map for the fmap with the worst RMSE among the displayed ones. It reveals that the measured fmaps strongly resemble the ideal ones and properly capture the image features. Errors are barely noticeable with the naked eye and consist of slightly different values between fmaps.\\
\indent Then, we turn to the assessment of the throughput and EE of the MAC operations. We first introduce the throughput as
\begin{equation}
	\textrm{Throughput} = \textrm{fps} \cdot N_\mathrm{filt} \cdot N_\mathrm{f}^2 \cdot (2 \cdot \textrm{F}^2 \cdot \textrm{DS}^2)\label{eq:throughput}
\end{equation}
where $N_\mathrm{filt}$ corresponds to the number of filters. This definition of the throughput does not account for the resolution of the inputs and weights involved in the MAC operations. Next, we can define the energy per 1b operation as
\begin{equation}
	\textrm{Energy/OP} = \frac{\textrm{Power}}{\textrm{fps} \cdot N_\mathrm{filt} \cdot N_\mathrm{f}^2 \cdot (2 \cdot \textrm{F}^2 \cdot \textrm{DS}^2) \cdot (B_\mathrm{X} \cdot B_\mathrm{W})}\label{eq:energy_per_1b_op}
\end{equation}
where $B_\mathrm{X}$ and $B_\mathrm{W}$ respectively stand for the resolution of the inputs and weights. In the proposed SoC, MAC operations are based on analog inputs and 4b weights. Hence, we use $B_\mathrm{X}$ = 1 and $B_\mathrm{W}$ = 4, even though using $B_\mathrm{X}$ equal to the effective number of bits (ENOB) at the input of the MAC units could be possible to compare the results with accelerators such as IMC ones, for which the resolution of inputs is clearly defined. Throughput can also be normalized to 1b operations by multiplying its expression in (\ref{eq:throughput}) by $B_\mathrm{X} \cdot B_\mathrm{W}$.\\
\indent Fig.~\ref{fig:19_meas_conv_mode_exposure}(a) illustrates different cases regarding how the exposure and convolution operations intertwine. A sequential execution is inefficient as pixels can start being exposed as soon as they have been stored in the analog memory. Therefore, a parallel execution is preferable. The current version of the imager controller only supports the case in which the exposure time $T_\mathrm{exp}$ is longer than the duration of the convolution operation $T_\mathrm{conv}$ (case 2), but could easily be modified to support a parallel execution for $T_\mathrm{exp} < T_\mathrm{conv}$ (case 3). This modification would be beneficial from an applicative standpoint, as it would allow to maximize the frame rate in all the configurations of the accelerator. Figs.~\ref{fig:19_meas_conv_mode_exposure}(b) and (c) correspond to an execution with four filters and a \mbox{12.5-ms} exposure time for all possible configurations of DS and S. Fig.~\ref{fig:19_meas_conv_mode_exposure}(b) reveals that a higher frame rate can be achieved with parallel execution, the limit being the exposure time.
\begin{figure}[!t]
	\centering
	\includegraphics[width=.4875\textwidth]{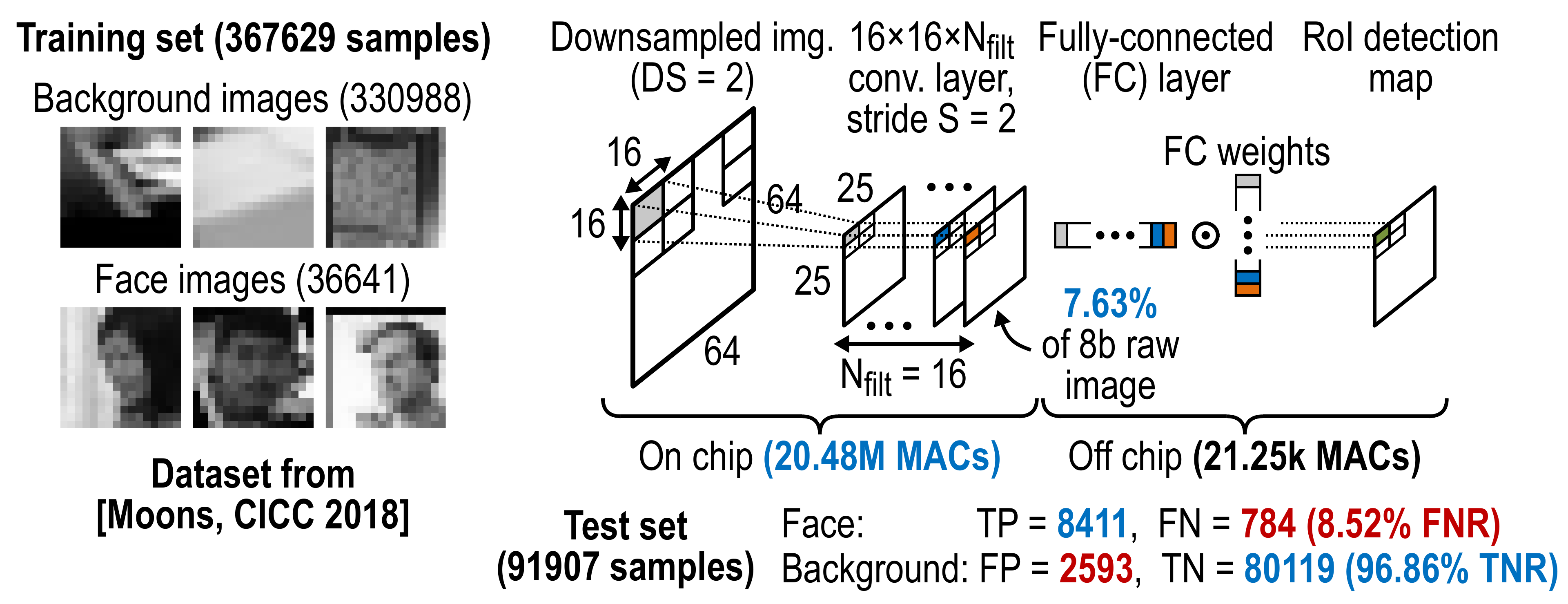}
	\caption{Training pipeline of the face RoI detector, based on a quantized CNN trained with QKeras and a TensorFlow backend (ideal software execution). The frame rate is 27~fps and is dominated by the duration of the convolution operation rather than by the exposure time.}
	\label{fig:22_meas_roi_detection_training}
\end{figure}
In terms of energy/OP at the SoC level, the parallel execution yields a reduction of 12 to 44$\%$ with the strongest reductions attained for small values of DS and S.\\
\indent Besides, Fig.~\ref{fig:20_meas_conv_mode_power} presents breakdowns of the power consumption and energy/OP at the levels of the accelerator and of the SoC. Regarding the SoC power [Fig.~ \ref{fig:20_meas_conv_mode_power}(a)], it ranges from 245 to 379~$\mu$W. The CPU and imager controller have a relatively constant consumption around 0.2~mW across configurations, while the consumption related to the analog circuitry and data transfers is highly dependent on the configuration [Fig.~\ref{fig:20_meas_conv_mode_power}(b)].
The power on $V_\mathrm{DDAL}$ and of the DMA declines for a larger DS and/or S, as $N_\mathrm{f}$ becomes smaller, while the power on $V_\mathrm{DDAH}$ and of the DCMI remains fairly constant. The former is indeed related to the pixel readout and DS3 units, and does not change as the frame rate is the same for all configurations except for DS = 1 and S = 2. This is the case because a parallel execution is employed and the frame rate is limited by the exposure time as in Fig.~\ref{fig:19_meas_conv_mode_exposure}(a). The latter is related to internal switching of the DCMI and does not account for the I/O power which would otherwise scale with the amount of data, similarly to the DMA. Therefore, the energy/OP at the SoC level [Fig.~\ref{fig:20_meas_conv_mode_power}(c)] goes from 0.22 to 6.43~pJ and degrades for large strides as the power is amortized over a smaller number of operations. Finally, the energy/OP at the accelerator level [Fig.~\ref{fig:20_meas_conv_mode_power}(d)] is comprised between 12 and 201~fJ, corresponding to an EE of 84.09 and 4.98~TOPS/W. Fig.~\ref{fig:20_meas_conv_mode_power}(d) and Table~\ref{table:summary_measured_performance} further reveal that the energy/OP at the accelerator level improves with a larger DS, as the filter is applied to a larger number of pixels in the original image thanks to the DS operation. Interestingly, the two key features to achieve a high EE at the accelerator level are the power gating of OTAs and the amortization of the MAC operation energy over a large number of pixels thanks to image DS.\\
\indent Finally, Fig.~\ref{fig:21_meas_conv_mode_nfilts} studies the impact of the number of filters $N_\mathrm{filt}$ on the frame rate, EE and energy/OP, for sequential exposure and convolution. Fig.~\ref{fig:21_meas_conv_mode_nfilts}(a) highlights that increasing $N_\mathrm{filt}$ causes the frame rate to drop as $T_\mathrm{conv}$ becomes longer while $T_\mathrm{exp}$ remains constant, and that the accelerator EE remains relatively constant while the SoC one slightly improves, as the fraction of time without convolution operations decreases and the digital power is amortized over a larger number of operations. The same trend is reflected by the energy/OP at the SoC level in Fig.~\ref{fig:21_meas_conv_mode_nfilts}(b).
\setlength{\tabcolsep}{2pt}
\begin{table*}[!t]
\centering
\caption{Comparison table of state-of-the-art mixed-signal vision chips.}
\label{table:soa_vision_chips}
\vspace{-0.15cm}
\scalebox{.8}{%
\begin{threeparttable}
\begin{scriptsize}
\begin{tabular}[t]{l|ccc|cc|cccc|c}
\toprule
& \multicolumn{3}{c}{\textbf{In-sensor}} & \multicolumn{2}{c}{\textbf{Hybrid}} & \multicolumn{5}{c}{\textbf{Near-sensor}}\\
\cmidrule(lr){2-4} \cmidrule(lr){5-6} \cmidrule(lr){7-11}
& Jendernalik & Carey & Xu & Lefebvre & Song & Kim / Bong & Young & Hsu & Hsu & \textbf{Lefebvre}\\
& \cite{Jendernalik_2013} & \cite{Carey_2013} & \cite{Xu_2022} & \cite{Lefebvre_2021} & \cite{Song_2021} & \cite{Kim_2017} / \cite{Bong_2018} & \cite{Young_2019} & \cite{Hsu_2021} & \cite{Hsu_2023} & \textbf{This work}\\
\midrule
Publication & TCAS-I & VLSI & TCAS-I & ISSCC & VLSI & ESSCIRC / JSSC & JSSC & JSSC & JSSC & JSSC\\
Year & 2013 & 2013 & 2022 & 2021 & 2021 & 2017 / 2018 & 2019 & 2021 & 2023 & 2024\\
\midrule
Technology & 0.35$\mu$m CMOS & 0.18$\mu$m CMOS & 0.18$\mu$m CMOS & 65nm CMOS & 0.18$\mu$m CIS & 65nm CMOS & 0.13$\mu$m CIS & 0.18$\mu$m CMOS & 0.18$\mu$m CMOS & 0.11$\mu$m CMOS\\
Area [mm$^2$] & 9.8 & 10$\times$10 & N/A & 2$\times$2 & 4.1$\times$5.2 & 3.3$\times$3.6 & 4$\times$4 & 2.46$\times$2 & 2.46$\times$2.18 & 1.37$\times$2.18\\
Supply voltage [V] & 3.3 & 1.8 (Digital) & 0.8--1.8 & 0.8/1 (Digital) & 1.8 (Digital) & 0.5--0.8 (Digital) & 0.9 (Digital) & 0.5 & 0.8 & 1.2 (Digital)\\
 & & 1.5 (Analog) & & 0.95/1.05 (Analog) & 2.5 (Analog) & 2.5 (Analog) & 1.5/2.5 (Analog) & & & 1.2/2.5/3.3 (Analog)\\
\midrule
Resolution & 64$\times$64 & 256$\times$256 & 32$\times$32 & 160$\times$128 & 240$\times$240 & 320$\times$240 & 320$\times$240 & 128$\times$128 & 126$\times$126 & 128$\times$128\\
Shutter & Global & Global & Global & Rolling & Global & Rolling & Global & Rolling & Rolling & Rolling\\
Double sampling & \textcolor{ECS-Red}{\textbf{No}} & \textcolor{ECS-Red}{\textbf{No}} & \textcolor{ECS-Blue}{\textbf{DRS}} & \textcolor{ECS-Red}{\textbf{No}} & \textcolor{ECS-Blue}{\textbf{CDS}} & \textcolor{ECS-Red}{\textbf{No}} & \textcolor{ECS-Blue}{\textbf{CDS}} & \textcolor{ECS-Red}{\textbf{No}} & \textcolor{ECS-Red}{\textbf{No}} & \textcolor{ECS-Blue}{\textbf{DRS}}\\
Frame rate [fps] & 100 & 100,000 & 156 & 24--268 & 120 & 1 & 30 & 480 & 50--250 & 18.2--79.7\\
Pixel pitch [$\mu$m] & \textcolor{ECS-Red}{\textbf{35}} & \textcolor{ECS-Red}{\textbf{32.3}} & \textcolor{ECS-Red}{\textbf{35}} & 9 & 9.8 & 7 & \textcolor{ECS-Blue}{\textbf{4}} & 7.6 & 7.6 & \textcolor{ECS-Blue}{\textbf{6.03}}\\
Pixel complexity & 18T APS + & 176T APS & 61T APS & 40T log(I) + & 14T APS + & 3T APS & 4T APS & 4T PWM & 4T PWM & 3T APS\\
 & 2 MOS caps. & & & 1 MIM cap. & 4 caps. & & & & & \\
Fill factor [$\%$] & 23 & \textcolor{ECS-Red}{\textbf{6.2}} & \textcolor{ECS-Red}{\textbf{9.1}} & 12.9 & 20.1 & N/A & \textcolor{ECS-Blue}{\textbf{60.4}} & 36 & 36 & \textcolor{ECS-Blue}{\textbf{54}}\\
DR [dB] & 58 & N/A & N/A & 47.1 & N/A & N/A & 59.3 & 52.3 & 47.8 & 57.7\\
\midrule
Feature type & - 3$\times$3 kernels & - Edge detection & - 32$\times$32 kernels & - 2$\times$2 to 64$\times$64 kernels & - Log. Haar filters & - 20$\times$20 lin. Haar filters & - Log. gradients & - 3$\times$3 kernels & - 3$\times$3 kernels & - 16$\times$16 kernels \\
 & & - Median filtering & & - 16$\times$16 lin. Haar filters & & & & & & \\
Multiscale & \textcolor{ECS-Red}{\textbf{No}} & \textcolor{ECS-Blue}{\textbf{Arbitrary}} & \textcolor{ECS-Red}{\textbf{No}} & - 6 scales (conv.) & \textcolor{ECS-Blue}{\textbf{Arbitrary}} & 3 scales & \textcolor{ECS-Blue}{\textbf{Arbitrary}} & \textcolor{ECS-Red}{\textbf{No}} & \textcolor{ECS-Red}{\textbf{No}} & 3 scales\\
 & & & & - 3 scales (Haar) & & & & & & \\
Computation type & Current & Current & Current & Current & Charge & Charge & Charge & Current & Current & Charge\\
Weight resolution & Analog & N/A & 1b & 1.5b & 1.5b & 1.5b & N/A & 4b & 4b & 4b\\
Feature resolution & Analog & 1b or 8b & 1b & 1b or 8b & 1b & 1b & 1.5b or 2.75b & 1b to 8b & 1b to 8b & 1b, 2b, 4b, or 8b\\
\midrule
Throughput\tnote{$\triangleleft$}\quad[MOPS] & \textcolor{ECS-Red}{\textbf{7.4}} & \textcolor{ECS-Blue}{\textbf{655,000}} & \textcolor{ECS-Red}{\textbf{5.1}} & 15.1--\textcolor{ECS-Blue}{\textbf{252.1}} & N/A & N/A & N/A & 137.2 & 63.5 & 10.5--\textcolor{ECS-Blue}{\textbf{408.3}}\\
Throughput\tnote{$\triangleleft \ddagger$}\quad[MOPS] & N/A & N/A & \textcolor{ECS-Red}{\textbf{5.1}} & 22.7--378.2 & N/A & N/A & N/A & \textcolor{ECS-Blue}{\textbf{548.7}} & 254 & 42--\textcolor{ECS-Blue}{\textbf{1633.2}}\\
Power (accel.) [$\mu$W] & N/A & N/A & 0.147--0.537 & N/A & N/A & N/A & N/A & N/A & N/A & 2.7--76.2\\
EE\tnote{$\ddagger$}\quad(accel.) [TOPS/W] & N/A & N/A & 9.52--\textcolor{ECS-Blue}{\textbf{34.77}} & N/A & N/A & N/A & N/A & N/A & N/A & 4.98--\textcolor{ECS-Blue}{\textbf{84.09}}\\
Power (SoC) [$\mu$W] & 280 & 1,230,000 & 8.5 & 42--206 & 2,900 & 24--96 & 229--262 & 117 & 80.4--134.5 & 250.9--384.7\\
EE\tnote{$\ddagger$}\quad(SoC) [TOPS/W] & \textcolor{ECS-Red}{\textbf{0.026}} & 0.53 & 0.60 & 0.23--\textcolor{ECS-Blue}{\textbf{5.46}} & N/A & N/A & N/A & \textcolor{ECS-Blue}{\textbf{4.67}} & 0.63--1.89 & 0.16--\textcolor{ECS-Blue}{\textbf{4.57}}\\
Processing energy (SoC) & \multirow{2}{*}{\textcolor{ECS-Red}{\textbf{683.6}}} & \multirow{2}{*}{187.7} & \multirow{2}{*}{\textcolor{ECS-Blue}{\textbf{3.3}}} & \multirow{2}{*}{\textcolor{ECS-Blue}{\textbf{2.5}}--103.9\tnote{$\star$}} & \multirow{2}{*}{16.8\tnote{$\dagger$}} & \multirow{2}{*}{\textcolor{ECS-Blue}{\textbf{6.0}}--24.0\tnote{$\diamond$}} & \multirow{2}{*}{49.7--56.9\tnote{$\triangleright$}} & \multirow{2}{*}{14.9} & \multirow{2}{*}{\textcolor{ECS-Blue}{\textbf{4.2}}--12.7} & \multirow{2}{*}{48.0--\textcolor{ECS-Red}{\textbf{284.1}}\tnote{$\star$}}\\
\mbox{[pJ/(pix$\cdot$frame$\cdot$filt)]} & & & & & & & & & & \\
\bottomrule
\end{tabular}
\end{scriptsize}
\begin{footnotesize}
\begin{tablenotes}
	\item[$\star$] For 4 filters.
	\item[$\dagger$] For 25 filters.
	\item[$\diamond$] For 52 filters.
	\item[$\triangleright$] Horizontal and vertical gradients are considered as two filters.
	\item[$\triangleleft$] Not normalized to the resolution of inputs and weights.
	\item[$\ddagger$] Normalized to 1b operations.
\end{tablenotes}
\end{footnotesize}
\end{threeparttable}%
}
\end{table*}

\vspace{-0.25cm}
\subsection{Face Region-of-Interest Detection}
\begin{figure}[!t]
	\centering
	\includegraphics[width=.4875\textwidth]{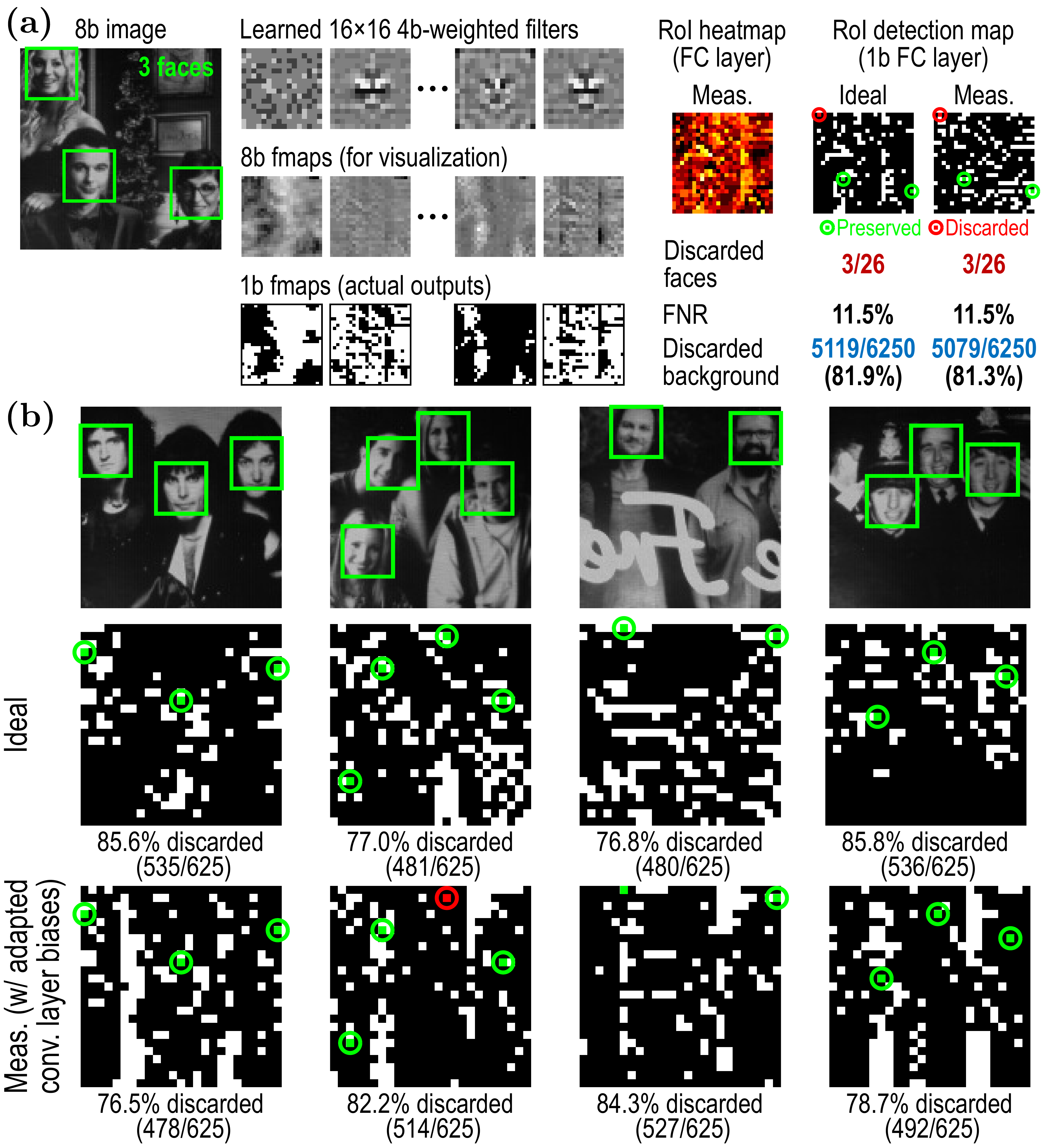}
	\caption{(a) Measured face RoI detection results, with details of a single test image and overall results for the 10 test images. (b) Face RoI results over four additional test images.}
	\label{fig:23_meas_roi_detection_results}
\end{figure}

This last experiment consists in demonstrating the operation of MANTIS in a face RoI detection use case. The structure and training pipeline of the RoI detector, implemented as a quantized CNN, are illustrated in Fig.~\ref{fig:22_meas_roi_detection_training}. The first part of the RoI detector is a convolution layer executed on chip, using 16 4b 16$\times$16 filters and 8b offsets, and operating over the image downsampled by 2$\times$. It is followed by an off-chip fully-connected (FC) layer with 8b weights, combining 1b fmaps to generate a 1b RoI detection map. Most of the workload is executed on chip, with 20.48 million operations in the convolution layer against 21.25 thousands in the FC one. Note that an ad-hoc digital accelerator could realize the FC layer on chip. An interesting feature of this detector is that it reduces the data that needs to be transmitted off chip to 7.63$\%$ of the raw 8b image, thus cutting down the I/O bandwidth by 13.1$\times$. As the EE at the SoC level for a sequential execution is 4.57~TOPS/W [Table~\ref{table:summary_measured_performance}] and the difference between the two execution types does not exceed 44$\%$ [Fig.~\ref{fig:19_meas_conv_mode_exposure}(c)], we expect the EE for a parallel execution to be above 2.56~TOPS/W. The network is trained with QKeras on a dataset consisting of background and face images used by Moons et al. in \cite{Moons_2018}, and achieves  false and true negative rates (FNR and TNR) of 8.5 and 96.9$\%$ on the test set, respectively, for an ideal software execution. At last, Fig.~\ref{fig:23_meas_roi_detection_results}(a) shows detailed results of one of the test images, and provides the overall performance over the 10 test images. MANTIS achieves an 11.5$\%$ FNR while respectively discarding 81.9$\%$ and 81.3$\%$ of image patches for the ideal and measured executions. These results are in line with the software execution but with a slight degradation coming from the fact that images generated by the imager are different from the ones in the dataset. Fig.~\ref{fig:23_meas_roi_detection_results}(b) displays face RoI results over four additional images, with a measured percentage of discarded image patches between 76.5 and 84.3$\%$, and a single discarded face. Interestingly, the overall performance remains largely similar whereas RoI detection maps are different for the ideal and measured executions, due to the adaptation of the biases of the convolution layer in measurement and an approximate modelling of raw pixel voltages' transformations inside the convolution pipeline.
\label{subsec:4C_face_RoI_detection}

\section{Comparison to the State of the Art}
\label{sec:5_comparison_to_the_state_of_the_art}
In this section, we compare our work to the state of the art of mixed-signal vision chips in Table~\ref{table:soa_vision_chips}, and to other relevant accelerators. The proposed SoC relies on a 3T APS with DRS to compensate FPN. This a key enabler to achieve a \mbox{6.03-$\mu$m} pixel pitch and a 54$\%$ fill factor, which are superior to existing vision chips, especially in-sensor ones for which the pixel pitch usually exceeds 30~$\mu$m. In terms of functionality, MANTIS is the first work to combine large 16$\times$16 filters, a 4b weight resolution, and operation at multiple scales, making it suitable for medium-complexity vision tasks such as RoI detection. The charge-domain MAC operations computed by the accelerator have an EE normalized to 1b operations (1b EE) between 4.98 and 84.09~TOPS/W, so 2.4$\times$ better than \cite{Xu_2022}. This information is however missing from other works which only report EE at the SoC level. The 1b EE at the SoC level ranges from 0.16 to 4.57~TOPS/W and is on par or better than existing works \cite{Hsu_2021, Lefebvre_2021}, while respectively supporting a larger filter size and an increased weight resolution. Regarding the processing energy, it is larger than for other works as this metric does not account for the filter characteristics. Next, let us compare near-sensor and hybrid vision chips. The latter usually present larger pixels which contain registers to store weights \cite{Lefebvre_2021}, or large capacitors to store and read pixel values \cite{Song_2021}. Regarding double sampling, it is possible when the readout is voltage-based \cite{Song_2021} but not when photocurrents are used \cite{Lefebvre_2021}. Lastly, near-sensor vision chips are generally more flexible as different configurations of the mixed-signal processor can easily be implemented. In contrast, hybrid architectures often involve hardwiring of connections between pixels, and either require to move weights in the pixel array \cite{Lefebvre_2021} or to combine row and column signals \cite{Song_2021}. These architectural features thus limit the use of hybrid vision chips to specific applications.\\
\indent Let us now consider other types of convolution accelerators, starting with those for which the digital control is implemented with an FPGA (not included in the power consumption). Two relevant works employ charge-domain computations, either using IMC with analog inputs and 1b weights \cite{Valavi_2019}, or in-column SC amplifiers with 3b weights, followed by ADCs performing a nonlinear quantization \cite{Jeong_2023}. Their 1b EE at the accelerator level is respectively 1.25 and 0.017~TOPS/W, which is approximately 4$\times$ lower than the worst-case EE obtained with MANTIS. Other works have their digital control embarked on chip. First, \cite{Abedin_2022} proposes a current-domain in-sensor processor with 2b weights based on resistive RAM, and attaining a 1b EE of 2.98~TOPS/W. Then, \cite{Wang_2024} features a hybrid optical-electronic CNN processor with 1b weights, realizing the first convolution layer with a mask in the optical domain, and reaching a 1b EE of 0.37~TOPS/W. Finally, \cite{Eki_2021} introduces a digital CNN processor stacked on an image sensor exhibiting a peak EE of 4.97~TOPS/W for 8b and 32b integer operations. MANTIS is on par with \cite{Abedin_2022, Wang_2024} in terms of EE, but offers more flexibility with its programmable convolution parameters. However, it is not as efficient as \cite{Eki_2021}, leveraging the performance improvements due to technology scaling.

\section{Conclusion}
\label{sec:6_conclusion}
In this work, we presented MANTIS, a mixed-signal near-sensor convolutional imager SoC intended for FE and RoI detection. It is the first mixed-signal vision chip to combine large 16$\times$16 filters with 4b weight resolution, operation at three different scales, and DRS to remove FPN and improve computational accuracy. MANTIS is enabled by two main circuit innovations. First, DS3 units combining DRS, voltage downshifting, and image DS. Next, near-sensor MAC operations computed in the charge domain, using SC amplifiers to compute psums and charge sharing in the capacitive DAC of the SAR ADCs to aggregate psums and compute the convolution result. MANTIS respectively reaches peak EEs normalized to 1b operations of 4.57 and 84.09~TOPS/W at the accelerator and SoC levels, while producing fmaps with an RMSE between 3.01 and 11.34$\%$. Lastly, face RoI detection was demonstrated with an FNR of 11.5$\%$, while discarding 81.3$\%$ of image patches and reducing the imager output data to 7.63$\%$ of the raw image. This work demonstrates that near-sensor vision chips can successfully tackle tasks requiring a higher resolution of inputs and weights, as opposed to in-sensor vision chips which are currently limited to noisy inputs and low-resolution weights. Further works should focus on the digital part of the SoC, albeit some analog blocks could also benefit from the utilization of more advanced techniques. New opportunities for the implementation of mixed-signal vision chips also arise from the advent of 2.5D/3D packaging.

% or


%\appendices
%\section{}
% Appendix text goes here.

\section*{Acknowledgments}
The authors would like to thank Prof. Marian Verhelst and Dr. Bert Moons for granting us access to their face detection dataset, Dr. R\'emi Dekimpe for his help with the DMA, Prof. Charlotte Frenkel and Dr. Adrian Kneip for fruitful discussions, and El\'eonore Masarweh for the chip microphotograph.

\ifCLASSOPTIONcaptionsoff
  \newpage
\fi



%\bibliographystyle{IEEEtran}
\bibliographystyle{IEEEtran}
\bibliography{Lefebvre_JSSC_2024_MANTIS}

\begin{thebibliography}{10}
\providecommand{\url}[1]{#1}
\csname url@samestyle\endcsname
\providecommand{\newblock}{\relax}
\providecommand{\bibinfo}[2]{#2}
\providecommand{\BIBentrySTDinterwordspacing}{\spaceskip=0pt\relax}
\providecommand{\BIBentryALTinterwordstretchfactor}{4}
\providecommand{\BIBentryALTinterwordspacing}{\spaceskip=\fontdimen2\font plus
\BIBentryALTinterwordstretchfactor\fontdimen3\font minus
  \fontdimen4\font\relax}
\providecommand{\BIBforeignlanguage}[2]{{%
\expandafter\ifx\csname l@#1\endcsname\relax
\typeout{** WARNING: IEEEtran.bst: No hyphenation pattern has been}%
\typeout{** loaded for the language `#1'. Using the pattern for}%
\typeout{** the default language instead.}%
\else
\language=\csname l@#1\endcsname
\fi
#2}}
\providecommand{\BIBdecl}{\relax}
\BIBdecl

\bibitem{Jendernalik_2013}
W.~Jendernalik, G.~Blakiewicz, J.~Jakusz, S.~Szczepanski, and R.~Piotrowski,
  ``{An Analog Sub-Milliwatt CMOS Image Sensor with Pixel-Level Convolution
  Processing},'' \emph{IEEE Trans. Circuits Syst. I, Reg. Papers}, vol.~60,
  no.~2, pp. 279--289, Feb. 2013.

\bibitem{Carey_2013}
S.~J. Carey, A.~Lopich, D.~R. Barr, B.~Wang, and P.~Dudek, ``{A 100,000 fps
  Vision Sensor with Embedded 535GOPS/W 256$\times$256 SIMD Processor Array},''
  in \emph{2013 Symp. VLSI Circuits}.\hskip 1em plus 0.5em minus 0.4em\relax
  IEEE, 2013, pp. C182--C183.

\bibitem{Xu_2022}
H.~Xu, N.~Lin, L.~Luo, Q.~Wei, R.~Wang, C.~Zhuo, X.~Yin, F.~Qiao, and H.~Yang,
  ``{Senputing: An Ultra-Low-Power Always-On Vision Perception Chip Featuring
  the Deep Fusion of Sensing and Computing},'' \emph{IEEE Trans. Circuits Syst.
  I, Reg. Papers}, vol.~69, no.~1, pp. 232--243, Jan. 2022.

\bibitem{Kim_2017}
C.~Kim, K.~Bong, I.~Hong, K.~Lee, S.~Choi, and H.-J. Yoo, ``{An Ultra-Low-Power
  and Mixed-Mode Event-Driven Face Detection SoC for Always-On Mobile
  Applications},'' in \emph{IEEE 43rd Eur. Solid-State Circuits Conf.}\hskip
  1em plus 0.5em minus 0.4em\relax IEEE, 2017, pp. 255--258.

\bibitem{Bong_2018}
K.~Bong, S.~Choi, C.~Kim, D.~Han, and H.-J. Yoo, ``{A Low-Power Convolutional
  Neural Network Face Recognition Processor and a CIS Integrated with Always-On
  Face Detector},'' \emph{IEEE J. Solid-State Circuits}, vol.~53, no.~1, pp.
  115--123, Jan. 2018.

\bibitem{Young_2019}
C.~Young, A.~Omid-Zohoor, P.~Lajevardi, and B.~Murmann, ``{A Data-Compressive
  1.5/2.75-bit Log-Gradient QVGA Image Sensor with Multi-Scale Readout for
  Always-On Object Detection},'' \emph{IEEE J. Solid-State Circuits}, vol.~54,
  no.~11, pp. 2932--2946, Nov. 2019.

\bibitem{Hsu_2021}
T.-H. Hsu, Y.-R. Chen, R.-S. Liu, C.-C. Lo, K.-T. Tang, M.-F. Chang, and C.-C.
  Hsieh, ``{A 0.5-V Real-Time Computational CMOS Image Sensor with Programmable
  Kernel for Feature Extraction},'' \emph{IEEE J. Solid-State Circuits},
  vol.~56, no.~5, pp. 1588--1596, May 2021.

\bibitem{Hsu_2023}
T.-H. Hsu, G.-C. Chen, Y.-R. Chen, R.-S. Liu, C.-C. Lo, K.-T. Tang, M.-F.
  Chang, and C.-C. Hsieh, ``{A 0.8 V Intelligent Vision Sensor with Tiny
  Convolutional Neural Network and Programmable Weights using Mixed-Mode
  Processing-in-Sensor Technique for Image Classification},'' \emph{IEEE J.
  Solid-State Circuits}, vol.~58, no.~11, pp. 3266--3274, Nov. 2023.

\bibitem{Lefebvre_2021}
M.~Lefebvre, L.~Moreau, R.~Dekimpe, and D.~Bol, ``{7.7 A 0.2-to-3.6 TOPS/W
  Programmable Convolutional Imager SoC with In-Sensor Current-Domain
  Ternary-Weighted MAC Operations for Feature Extraction and Region-of-Interest
  Detection},'' in \emph{2021 IEEE Int. Solid-State Circuits Conf.},
  vol.~64.\hskip 1em plus 0.5em minus 0.4em\relax IEEE, 2021, pp. 118--120.

\bibitem{Song_2021}
H.~Song, S.~Oh, J.~Salinas, S.-Y. Park, and E.~Yoon, ``{A 5.1 ms Low-Latency
  Face Detection Imager with In-Memory Charge-Domain Computing of
  Machine-Learning Classifiers},'' in \emph{2021 Symp. VLSI Circuits}.\hskip
  1em plus 0.5em minus 0.4em\relax IEEE, 2021, pp. 1--2.

\bibitem{Lefebvre_2024}
M.~Lefebvre and D.~Bol, ``{A Mixed-Signal Near-Sensor Convolutional Imager SoC
  with Charge-Based 4b-Weighted 5-to-84-TOPS/W MAC Operations for Feature
  Extraction and Region-of-Interest Detection},'' in \emph{2024 IEEE Custom
  Integr. Circuits Conf.}\hskip 1em plus 0.5em minus 0.4em\relax IEEE, 2024,
  pp. 1--2.

\bibitem{Gönen_2017}
B.~G{\"o}nen, F.~Sebastiano, R.~Quan, R.~van Veldhoven, and K.~A. Makinwa, ``{A
  Dynamic Zoom ADC with 109-dB DR for Audio Applications},'' \emph{IEEE J.
  Solid-State Circuits}, vol.~52, no.~6, pp. 1542--1550, June 2017.

\bibitem{Park_2020}
I.~Park, W.~Jo, C.~Park, B.~Park, J.~Cheon, and Y.~Chae, ``{A 640 $\times$ 640
  Fully Dynamic CMOS Image Sensor for Always-On Operation},'' \emph{IEEE J.
  Solid-State Circuits}, vol.~55, no.~4, pp. 898--907, Apr. 2020.

\bibitem{Kneip_2023}
A.~Kneip, M.~Lefebvre, J.~Verecken, and D.~Bol, ``{IMPACT: A 1-to-4b 813-TOPS/W
  22-nm FD-SOI Compute-in-Memory CNN Accelerator Featuring a 4.2-POPS/W
  146-TOPS/mm 2 CIM-SRAM With Multi-Bit Analog Batch-Normalization},''
  \emph{IEEE J. Solid-State Circuits}, vol.~58, no.~7, pp. 1871--1884, May
  2023.

\bibitem{Seo_2023}
J.-O. Seo, M.~Seok, and S.~Cho, ``{A 44.2-TOPS/W CNN Processor With
  Variation-Tolerant Analog Datapath and Variation Compensating Circuit},''
  \emph{IEEE J. Solid-State Circuits}, vol.~59, no.~5, May 2023.

\bibitem{Ginsburg_2005}
B.~P. Ginsburg and A.~P. Chandrakasan, ``{An Energy-Efficient Charge Recycling
  Approach for a SAR Converter with Capacitive DAC},'' in \emph{2005 IEEE Int.
  Symp. Circuits Syst.}\hskip 1em plus 0.5em minus 0.4em\relax IEEE, 2005, pp.
  184--187.

\bibitem{Li_2014}
Y.~Li and Y.~Lian, ``{Improved Binary-Weighted Split-Capacitive-Array DAC for
  High-Resolution SAR ADCs},'' \emph{Electron. Lett.}, vol.~50, no.~17, pp.
  1194--1195, Aug. 2014.

\bibitem{Harpe_2019}
P.~Harpe, ``{A Compact 10-b SAR ADC with Unit-Length Capacitors and a Passive
  FIR Filter},'' \emph{IEEE J. Solid-State Circuits}, vol.~54, no.~3, pp.
  636--645, Mar. 2019.

\bibitem{Shanbhag_2022}
N.~R. Shanbhag and S.~K. Roy, ``{Benchmarking In-Memory Computing
  Architectures},'' \emph{IEEE Open J. Solid-State Circuits Soc.}, vol.~2, pp.
  288--300, Dec. 2022.

\bibitem{Moons_2018}
B.~Moons, D.~Bankman, L.~Yang, B.~Murmann, and M.~Verhelst, ``{BinarEye: An
  Always-On Energy-Accuracy-Scalable Binary CNN Processor with All Memory on
  Chip in 28nm CMOS},'' in \emph{2018 IEEE Custom Integr. Circuits Conf.}\hskip
  1em plus 0.5em minus 0.4em\relax IEEE, 2018, pp. 1--4.

\bibitem{Valavi_2019}
H.~Valavi, P.~J. Ramadge, E.~Nestler, and N.~Verma, ``{A 64-Tile 2.4-Mb
  In-Memory-Computing CNN Accelerator Employing Charge-Domain Compute},''
  \emph{IEEE J. Solid-State Circuits}, vol.~54, no.~6, pp. 1789--1799, June
  2019.

\bibitem{Jeong_2023}
B.~Jeong, J.~Lee, J.~Choi, M.~Song, Y.~Son, and S.~Y. Kim, ``{A 0.57 mW@1 FPS
  In-Column Analog CNN Processor Integrated Into CMOS Image Sensor},''
  \emph{IEEE Access}, vol.~11, June 2023.

\bibitem{Abedin_2022}
M.~Abedin, A.~Roohi, M.~Liehr, N.~Cady, and S.~Angizi, ``{MR-PIPA: An
  Integrated Multilevel RRAM (HfO$_\mathrm{x}$)-Based Processing-in-Pixel
  Accelerator},'' \emph{IEEE J. Explor. Solid-State Comput. Devices Circuits},
  vol.~8, no.~2, pp. 59--67, Sept. 2022.

\bibitem{Wang_2024}
X.~Wang, Z.~Huang, T.~Liu, W.~Shi, H.~Chen, and M.~Zhang, ``{6.9 A 0.35 V 0.367
  TOPS/W Image Sensor with 3-Layer Optical-Electronic Hybrid Convolutional
  Neural Network},'' in \emph{2024 IEEE Int. Solid-State Circuits Conf.},
  vol.~67.\hskip 1em plus 0.5em minus 0.4em\relax IEEE, 2024, pp. 116--118.

\bibitem{Eki_2021}
R.~Eki, S.~Yamada, H.~Ozawa, H.~Kai, K.~Okuike, H.~Gowtham, H.~Nakanishi,
  E.~Almog, Y.~Livne, G.~Yuval \emph{et~al.}, ``{9.6 A 1/2.3inch 12.3Mpixel
  with On-Chip 4.97 TOPS/W CNN Processor Back-Illuminated Stacked CMOS Image
  Sensor},'' in \emph{2021 IEEE Int. Solid-State Circuits Conf.},
  vol.~64.\hskip 1em plus 0.5em minus 0.4em\relax IEEE, 2021, pp. 154--156.

\end{thebibliography}


% Martin Lefebvre
\vspace{-1cm}
\begin{IEEEbiography}[{\includegraphics[width=1in,height=1.25in,clip,keepaspectratio]{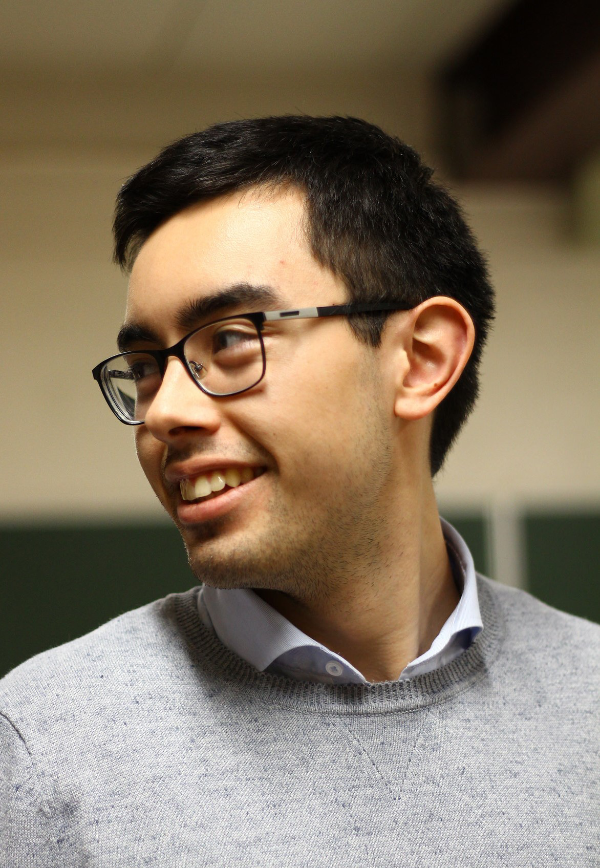}}]{Martin Lefebvre} (Graduate Student Member, IEEE) received the M.Sc. and Ph.D. degrees in engineering sciences from the Universit\'e catholique de Louvain (UCLouvain), Belgium, in 2017 and 2024.

He is currently a post-doctoral researcher with the cognitive sensor nodes and systems (CogSys) laboratory led by Prof. C. Frenkel at TU Delft, The Netherlands, working on neuromorphic hardware/software co-design for efficient on-chip learning. His research interests include hardware-aware machine learning algorithms, mixed-signal vision chips for embedded image processing, and low-power current reference architectures.

Dr. Lefebvre serves as a reviewer for various IEEE journals and conferences including \sc{IEEE Journal of Solid-State Circuits} and \sc{IEEE Transactions on Circuits and Systems I and II}.
\end{IEEEbiography}

% David Bol
\vspace{-1cm}
\begin{IEEEbiography}[{\includegraphics[width=1in,height=1.25in,clip,keepaspectratio]{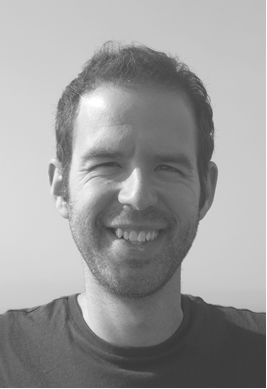}}]{David Bol} (Senior Member, IEEE) received the Ph.D. degree in engineering science from the Université catholique de Louvain (UCLouvain), in 2008, in the field of ultra-low-power digital nanoelectronics.

In 2005, he was a visiting Ph.D. student at the CNM, Sevilla, and in 2009, a post-doctoral researcher at intoPIX, Louvain-la-Neuve. In 2010, he was a visiting post-doctoral researcher at the UC Berkeley Lab for Manufacturing and Sustainability, Berkeley. In 2015, he participated to the creation of e-peas semiconductors spin-off company. Prof. Bol leads the Electronic Circuits and Systems (ECS) group, UCLouvain, focused on ultra-low-power design of integrated circuits for environmental and biomedical IoT applications including computing, power management, sensing and wireless communications. He is currently an Associate Professor with UCLouvain. He has authored more than 150 papers and conference contributions and holds three delivered patents. He is actively engaged in a social-ecological transition in the field of information and communication technologies (ICT) research with a post-growth approach.

Prof. Bol (co-)received five Best Paper/Poster/Design Awards in IEEE conferences (ICCD 2008, SOI Conf. 2008, FTFC 2014, ISCAS 2020, ESSCIRC 2022) and supervised the Ph.D. thesis of Charlotte Frenkel who received the 2021 Nokia Bell Scientific Award and the 2021 IBM Innovation Award for her Ph.D. He serves as a reviewer for various IEEE journals and conferences and presented several keynotes in international conferences. 
\end{IEEEbiography}


%\vfill


\end{document}